\documentclass[aps,prb,twocolumn,floats]{revtex4-2}
\usepackage{epsfig}
\usepackage{amsmath, amssymb}
\usepackage{graphicx}
\usepackage{soul}
\usepackage{braket}
\usepackage[colorlinks=true,linkcolor=blue]{hyperref}
\usepackage{subfigure}
\usepackage{color}

\newcommand{\blue}{\textcolor{blue}}

\begin{document}
%\raggedbottom
\title{Topology and $\mathcal{PT}$ Symmetry in a Non-Hermitian Su-Schrieffer-Heeger Chain with Periodic Hopping Modulation}
\author{Surajit Mandal$^{1,2}$}
\email{surajitmandalju@gmail.com}
\author{Satyaki Kar$^2$}
\email{satyaki.phys@gmail.com  (corresponding author)}
\affiliation{$^1$Department of Physics, Jadavpur University, Kolkata - 700032, West Bengal, India\\
$^2$Department of Physics, AKPC Mahavidyalaya, Bengai, West Bengal -712611, India}
%\vskip -3 in
%\twocolumn[
%\begin{@twocolumnfalse}
%  \maketitle
\begin{abstract}
  We study the effect of periodic {but commensurate} hopping modulation on a Su-Schrieffer-Heeger (SSH) chain with an additional onsite staggered imaginary potential. Such dissipative, non-Hermitian (NH) extension amply modifies the features of the topological trivial phase (TTP) and the topological nontrivial phase (TNP) of the SSH chain, {more so with the periodic hopping distribution. Generally} a weak potential can respect the parity-time ($\mathcal{PT}$) symmetry keeping the energy eigenvalues real, while a strong potential breaks $\mathcal{PT}$ conservation leading to imaginary end state and complex bulk state energies in the system. {We find that this $\mathcal{PT}$ breaking with imaginary potential strength $\gamma$ show interesting dependence on the hopping modulation $\Delta$ for different hoping modulations. In-gap states, that appear also in the $\gamma=0$ limit, take either purely real or purely imaginary eigenvalues depending on the strength of both $\gamma$ and $\Delta$. The localization of end states (in-gap states) at the boundaries are investigated which show extended nature not only near topological transitions (further away from $|\Delta/t|=1$) but also near the unmodulated limit of $\Delta=0$. Moreover, localization of the bulk states is observed at the maximally dimerized limit of $|\Delta/t|=1$, which also have a $\gamma$ dependence. Analyzing further the dissipation caused by the complex eigenvalues in this problem with different hopping periodicity can be essential in modulating the gain-loss contrast in optical systems or in designing various quantum information processing and storage devices. }
%     \keywords{}
\end{abstract}
%\date
%\end{@twocolumnfalse}
%]
\maketitle
%\vskip 1 in
\section{Introduction}\label{sec1}
As per the Dirac-Von Neumann's formulation in quantum mechanics, all the physical observables in the Hilbert space are represented by Hermitian operators\cite{shankar}, which in turn gives the real eigenvalues in the energy spectrum with the assurance of the conservation of probability. Interestingly, a wide class of NH Hamiltonians that respect $\mathcal{PT}$-symmetry can also display entirely real energy eigenvalues\cite{bender1,bender2} thereby garnering huge attention nowadays from the physics community.

{A Hermitian SSH chain can exist in two topologically distinct phases characterized by the presence or absence of zero energy end states and separated by a topological quantum phase transition (TQPT) occurring at the dimerization parameter $\Delta=0$. The ten-fold symmetry classes of such Hermitian systems can be ramified into much wider varieties due to the difference between transposition and complex conjugation, when a NH term/potential is introduced there\cite{prx9}.} Such NH extension features a $\mathcal{PT}$ transition that separates $\mathcal{PT}$-unbroken and $\mathcal{PT}$-broken phases with the NH Hamiltonian giving complex spectrum in the $\mathcal{PT}$-broken phase\cite{lieu}. Thus the non-Hermitian (NH) Hamiltonian, unlike its Hermitian counterpart, features band gap closure along real and/or imaginary energy axis. The end modes that appear in TNP in a finite chain, soon become imaginary on switching on the NH potential, though the bulk bands still remains gapless there along the imaginary energy axis yet being gapped along the real axis. {However for a chain with periodic boundary condition (PBC), end modes are absent and hence $\mathcal{PT}$ transition takes place with bulk mode eigenvalues starting to become complex for higher values of $\gamma$, be it either in the topological or trivial phase. Such transitions occur symmetrically away from the TQPT for a range of $\Delta$ the separation between the two being proportional to $\gamma$. As the bulk-boundary correspondance, a global invariant such as a complex Berry phase can be defined in a periodic chain that implies existence of topological end modes within TNP in a finite chain,} in compatible with the pseudo-anti-Hermiticity of a $\mathcal{PT}$-symmetric model\cite{edge2,lieu}. The bulk band crossings at $\mathcal{PT}$ transition (bulk modes turning real to complex conjugate pairs is referred to as crossing here\cite{lieu}) occur far from the topological transition point and thus the invariant remains a property of the entire Hamiltonian instead of individual bands\cite{lieu}.

{As already mentioned, the SSH chain with open boundary condition (OBC) exhibits topologically protected zero energy mid-gap states while its NH extension shows topological modes having complex energy eigenvalues\cite{lieu,edge2,wang,pt,zhu,slootman}. Such modes in a Hermitian system are localized at the edges while the bulk modes are of extended nature in general. Non-Hermiticity introduced via consideration of non-reciprocal lattices can cause the bulk states also to become skewed towards the edges causing a NH skin effect\cite{skin}. Though a diagonal NH potential don't show such features in general, we should remember that a periodic hopping modulation in a Hermitian SSH chain (in simple reciprocal lattices) creates non-topological in-gap states\cite{mandal,sk} which remain localized at edges. Its worth studying the effect of $\gamma$ on them or any other bulk or end states in a NH SSH chain.}

The intricate interplay between the topological properties and the spontaneously $\mathcal{PT}$ breaking transition ($\mathcal{SPT~BT}$) in a SSH {and Kitaev chain} has been studied previously by Klett $et.~al.$\cite{pt} and Wang $et.~al.$\cite{wang} respectively.
{On the other hand, a thorough spectral and topological analysis of a Hermitian SSH chain with periodically modulated hopping is presented in Ref.\cite{mandal,sk}.}
In this paper, we aim to extend such analysis in presence of additional staggered imaginary potentials which facilitates extraction of further novel phenomena in regards to topological and $\mathcal{PT}$-symmetric/broken phases. 
 {Within TNP in a finite chain, the advent of spontaneous $\mathcal{PT}$ braking occurs at a point called an exceptional point (EP) where the end state eigenvalues coalesces to zero\cite{ep}. We find that such points correspond to practically zero values of $\gamma$ away from TQPT point (within TNP). In TTP, however, spontaneous $\mathcal{PT}$ broken phases ($\mathcal{SPT~BP}$) appears when the bulk modes become complex in energy.} The appearance of $\mathcal{SPT~BP}$ in the TNP for these systems have similarities with a different extension of the SSH model where non-Hermiticity enters only via imaginary boundary potentials\cite{zhu}.

The paper is organized as follows. In Section II, formulation of our $\mathcal{PT}$ symmetric model is given. Section III discusses the numerical results for commensurate variation of hopping periodicity given by $\theta=\pi,~\pi/2$ and $\pi/4$ (to be defined later) and provides a comparative analysis among them. Finally, in Section IV, we summarize the findings to conclude our work and mention about the practicability of the problem and its future possibilities in the field of topological quantum computation.

\section{$\mathcal{PT}$-Symmetric Model}\label{sec2}
One dimensional Hamiltonian for NH SSH model with staggered nearest-neighbor hopping and onsite imaginary potentials becomes
\begin{equation}\label{1}
  \mathcal{H}_{\mathcal{PT}}=\mathcal{H}_{SSH}+U,
\end{equation}
in which $\mathcal{H}_{SSH}$ is the Hamiltonian of SSH chain of $L=M*N$ ($M$ and $N$ are the number sublattice and unit cells respectively) sites with modulated hopping strength and defined as 
\begin{equation}\label{2}
  \mathcal{H}_{SSH}=\sum_{i}^{L-1}(t+\delta_{i})[\ket{i}\bra{i+1}+H.c],
\end{equation}
where $\delta_{i}=\Delta\cos[(i-1)\theta]$ with $i=1,2,3,......, n$ gives the periodic modulation in hopping strength (t). Generally, one acquire $\delta_{i+1}=\Delta\cos(\frac{2\pi i}{M})$ when $\theta=\frac{2\pi}{M}$ and the chain shows a $M$ sublattice structure. One can represent the system by a $M\times M$ Hamiltonian matrix having $M$ number of eigenmodes. The onsite term $U$ is $\mathcal{PT}$-symmetric and is given by
\begin{equation}\label{3}
  U=i\gamma\sum_{i}^{L}(-1)^{i-1}\ket{i}\bra{i},
\end{equation}
in which the coefficient $\gamma~(>0)$ is called the gain-loss contrast as it indicates the gain or loss of particles in this dissipative 1D system. Notice that Eq.(\ref{2}) is chiral symmetric while Eq.(\ref{1}) is not. However, Eq.(\ref{1}) is $\mathcal{PT}$ symmetric. The effects of parity (space-reflection) operator, $\mathcal{P}$ and time-reversal operator, $\mathcal{T}$ are given by $p\rightarrow-p$,~$x\rightarrow-x$ and $p\rightarrow-p$,~$x\rightarrow x$,~$i\rightarrow-i$, respectively. For discrete systems, the effects of these two operators are defined as $\mathcal{P}c_{i}(c^{\dagger}_{i})\mathcal{P}=c_{L+1-i}(c^{\dagger}_{L+1-i})$ and $\mathcal{T}i\mathcal{T}=-i$. Though $\mathcal{P}$ and $\mathcal{T}$ do not individually commute with $\mathcal{H}_{\mathcal{PT}}$, the Hamiltonian remains invariant under their combined effect, $[\mathcal{\hat{PT}},~\mathcal{H}_{\mathcal{PT}}]=0$ (with $\mathcal{\hat{PT}}$ is the $\mathcal{PT}$ symmetry operator) which helps the complex energy eigenvalues to become entirely real in the $\mathcal{PT}$-symmetric eigenstates. However, there are parameter regimes where the Hamiltonian symmetry is broken spontaneously in the eigenstates thereby producing complex eigenvalues\cite{operator}.

It is obvious that chiral or sublattice symmetry which protects the end states in a SSH model can be given by the operator $\tau_z=\mathbb{I}_N\otimes\sigma_z$ with $\mathbb{I}_{N}$ referring to the $N\times N$ identity matrix\cite{ep}. Hence the single-particle SSH Hamiltonian\cite{ep} follows the anticommutation: $\{\mathcal{H}_{SSH},~\tau_{z}\}=0$. However, this no longer holds good for the NH model Hamiltonian Eq.(\ref{1}) in the presence of a staggered imaginary potential: $\gamma\ne 0$\cite{lieu} and the underlying symmetry relation needs to be modified for NH systems\cite{chiral}. From an analogy of the Hermitian case, here one can think of a symmetry operator $\Lambda=\tau_{z}\mathcal{T}$, in which the time-reversal operator $\mathcal{T}$ becomes merely a complex conjugation for this spin-polarized/spinless case\cite{operator,ep}. Like the chiral operator of the Hermitian case, with the operator $\Lambda$ one can have a generalized version of the Hamiltonian transformation by which the non-Hermitian satisfies $\Lambda^{\dagger}\mathcal{H}_{\mathcal{PT}}\Lambda=-\mathcal{H}_{PT}^\dagger$ indicating its ``pseudo-anti-Hermiticity" property\cite{operator}. The combined action of $\mathcal{P}$ and $\mathcal{T}$ makes the spectrum symmetric about both the real and imaginary axis and thus one can have a quartet of {$\mathcal{PT}$ symmetric states with energies $E,-E, E^{\star} \rm{and} -E^\star$}\cite{bender1,bender2,ep,symmetry,ruter,edge2}.

As per the symmetry of the eigenfunctions\cite{bender1,bender2}, the NH system can possess a broken or an unbroken $\mathcal{PT}$-symmetry. Recalling the time-independent Schr\"{o}dinger equation of eigenvectors $\ket{\psi}$ as
\begin{equation}\label{4}
\mathcal{H}_{\mathcal{PT}}\ket{\psi}=E\ket{\psi},
\end{equation}
\begin{widetext}
\begin{table*}
\parbox{.82\linewidth}{
\centering
\begin{tabular}{  p{1.5cm}| p{3cm}| p{5.5cm} |p{5.5cm} } 
 \hline
 $\theta$ values & TQPT point & TNP & TTP \\ [0.65ex] 
 \hline
 $\pi$ & $\Delta/t=0$ & $\Delta/t<0$ & $\Delta/t>0$ \\ [1.2ex] 
 \hline
  $\pi/2$ & $|\Delta/t|=\sqrt{2}$ & $0<|\Delta/t|<\sqrt{2}$ & $|\Delta/t|>\sqrt{2}$ \\ [1.2ex] 
 \hline
  $\pi/4$ & $|\Delta/t|=\sqrt{2(2\pm\sqrt{2})}$ & $0<|\Delta/t|<\sqrt{2(2-\sqrt{2})};\ |\Delta/t|>\sqrt{2(2+\sqrt{2})}$ & $\sqrt{2(2-\sqrt{2})}<|\Delta/t|<\sqrt{2(2+\sqrt{2})}$ \\[1.2ex] 
 \hline
\end{tabular}
\caption{TQPT point, domain of the topological and trivial phases of periodic modulated SSH chain for different $\theta$ values. This range of TNP and TTP is based on an earlier study as found in Ref.\cite{mandal}.}
\label{table:1}}
\end{table*}
\end{widetext} 
where $E$ is the eigenvalues. The system should acquire real eigenvalues and respect $\mathcal{PT}$-symmetry if all the eigenvectors follow $\mathcal{PT}$-symmetry relation $\mathcal{PT}\ket{\psi}=\ket{\psi}$. However, the system generates broken $\mathcal{PT}$-symmetry if all the eigenfunctions do not obey this relation, and the corresponding eigenvalues are complex\cite{zhu}.

In the absence of imaginary potential Eq.(\ref{3}), Eq.(\ref{1}) reduces to a hopping modulated Hermitian SSH chain\cite{mandal}. For $\theta=\pi$, this chain shows a two-band spectrum. The quantum phase transition point within the band spectra is important in finding the topological domain and such domain for various commensurate $\theta$ values for the isolated/Hermitian (when $\gamma=~0$) SSH chain is given in Table.\ref{table:1}. The zero energy states (ZES) found in the domain of topological phase for an isolated SSH chain (as mentioned in Table.\ref{table:1}) are protected by particle-hole as well as inversion symmetry\cite{ryu}. In the following section, we mainly focused on the TNP and TTP for all the commensurate $\theta$ values to see their exotic behavior in the presence of an alternating gain and loss strength $\gamma$ in terms of an onsite imaginary potential.
%{For the NH systems, the $\mathcal{PT}$ symmetric transition point is a crossover after that the system enters into a completely $\mathcal{PT}$-symmetry broken region\cite{lieu}.}

\section{Numerical Result}
This section is devoted to accumulating the numerical calculations of eigenvalue Eq.(\ref{4}) of the above mentioned models under the consideration of OBC. We delve into investigating the effects of alternating gain and loss on the topological/trivial regime of the energy spectrum for the concerned system.
\subsection{Case of $\theta=\pi$}
The considered model, for this $\theta$ value, is staggered by modulation parameter $\pm\Delta$ which in turn fully corroborates the usual SSH chain. In bulk momentum (k) space, the model Eq.(\ref{1}) gives the single-particle Bloch Hamiltonian for $\theta=\pi$ as:
\begin{equation}\label{5}
\mathcal{H}_{\mathcal{PT}}(k)=
\begin{pmatrix}
i\gamma & A_{k} \\
 A_{-k} & -i\gamma
\end{pmatrix},
\end{equation}
where, $A_{k}=(t+\Delta)+ (t-\Delta)e^{-ik}$. In terms of Pauli matrices $\sigma_{i}$'s this becomes
\begin{equation}\label{5a}
  \scriptsize
\mathcal{H}_{\mathcal{PT}}(k)=[(t+\Delta)+(t-\Delta)cos(k)]\sigma_{x}+(t-\Delta)sin(k)\sigma_{y}\\+i\gamma\sigma_{z}
\end{equation}
and the dispersion relation comes out to be
\begin{equation}\label{6}
E(k)_{\pm}=\pm\sqrt{2[t^2+\Delta^2+(t^2-\Delta^2)cosk]-\gamma^2}
\end{equation}
\begin{figure}
   \vskip -.4 in
   \begin{picture}(100,100)
     \put(-70,0){
   \includegraphics[width=.45\linewidth]{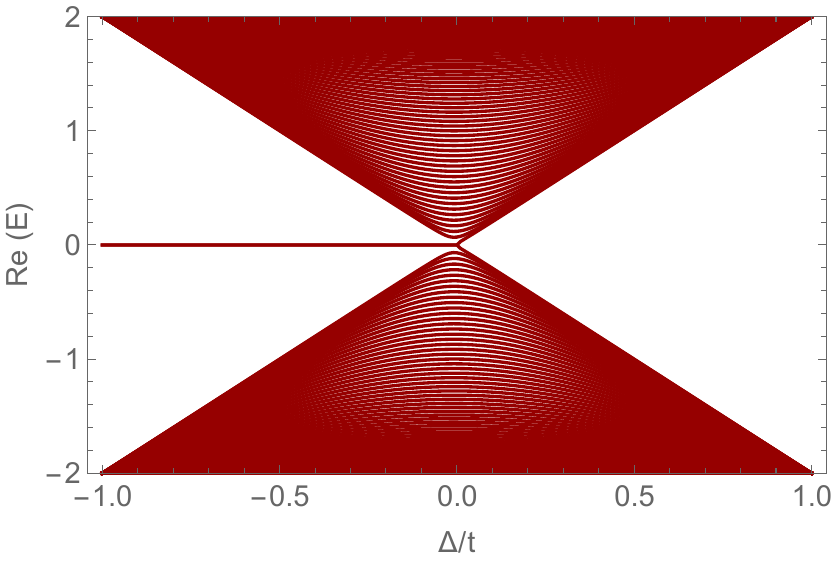}\hskip .2 in
   \includegraphics[width=.45\linewidth]{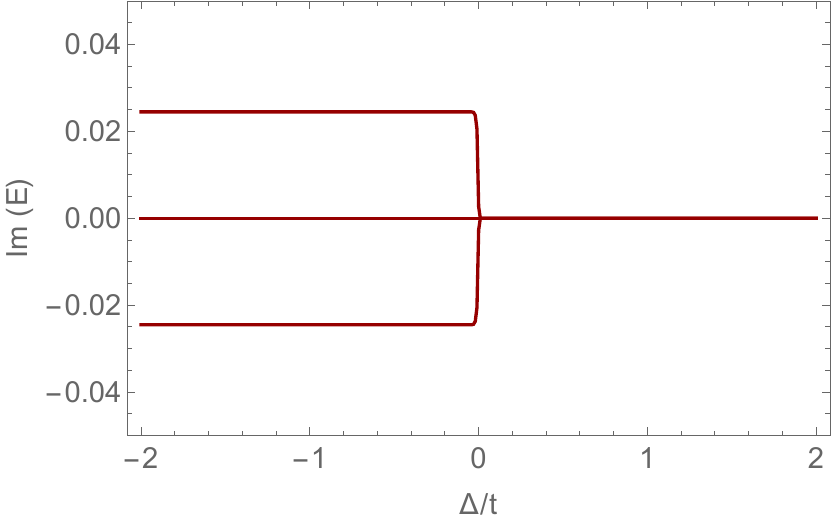}}
   \put(45,35){(a)}
   \end{picture}\\
    \vskip -.4 in
  \begin{picture}(100,100)
     \put(-70,0){
   \includegraphics[width=.45\linewidth]{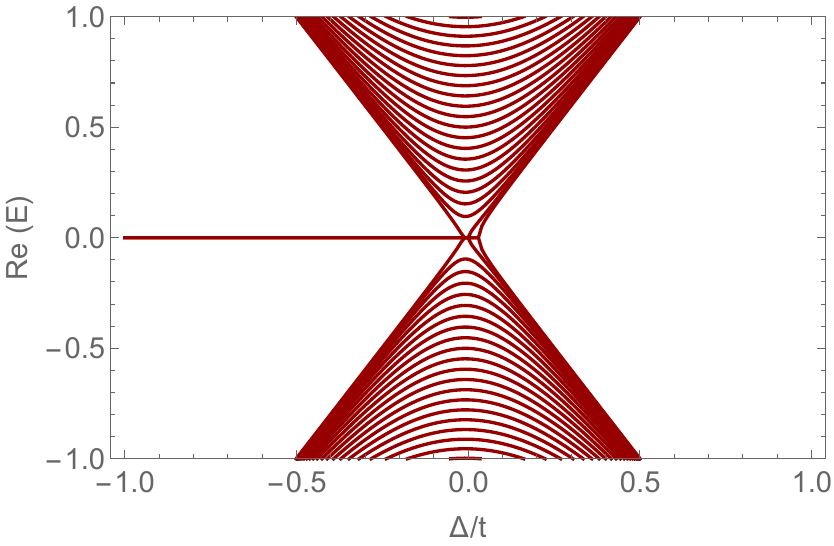}\hskip .2 in
   \includegraphics[width=.45\linewidth]{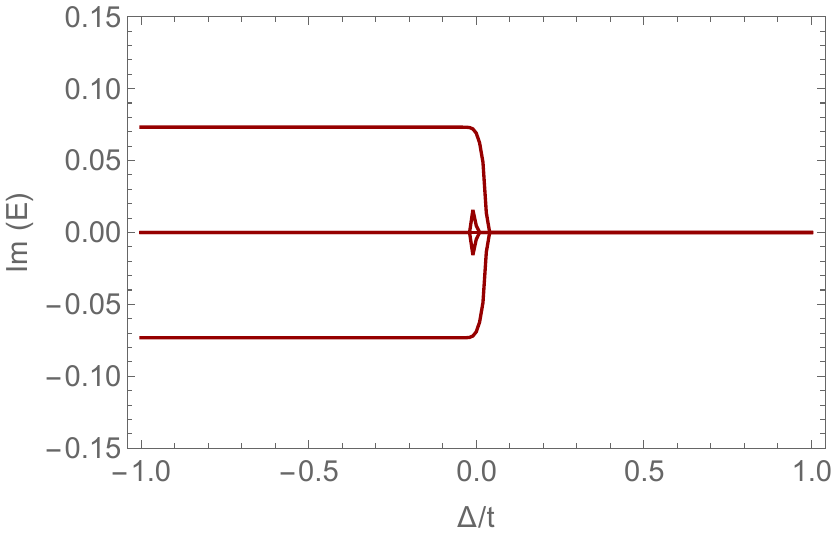}}
   \put(45,35){(b)}
   \end{picture}\\
    \vskip -.4 in \begin{picture}(100,100)
     \put(-70,0){
   \includegraphics[width=.45\linewidth]{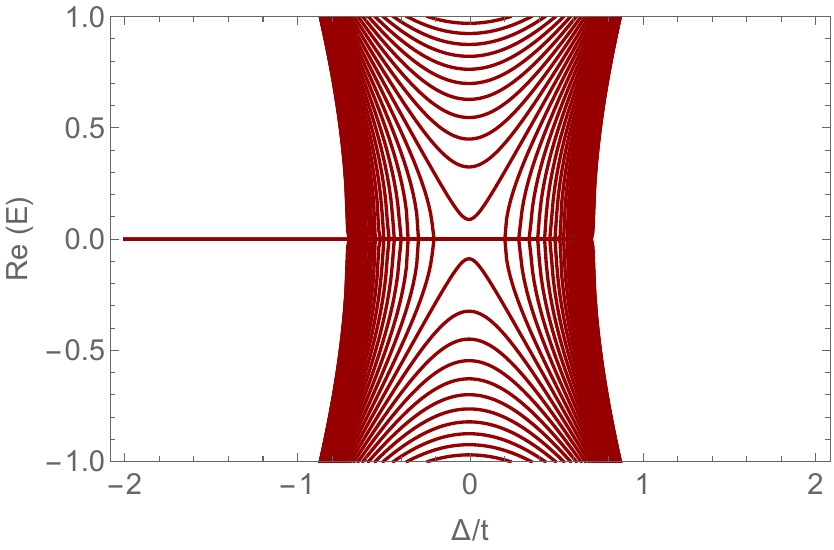}\hskip .2 in
   \includegraphics[width=.45\linewidth]{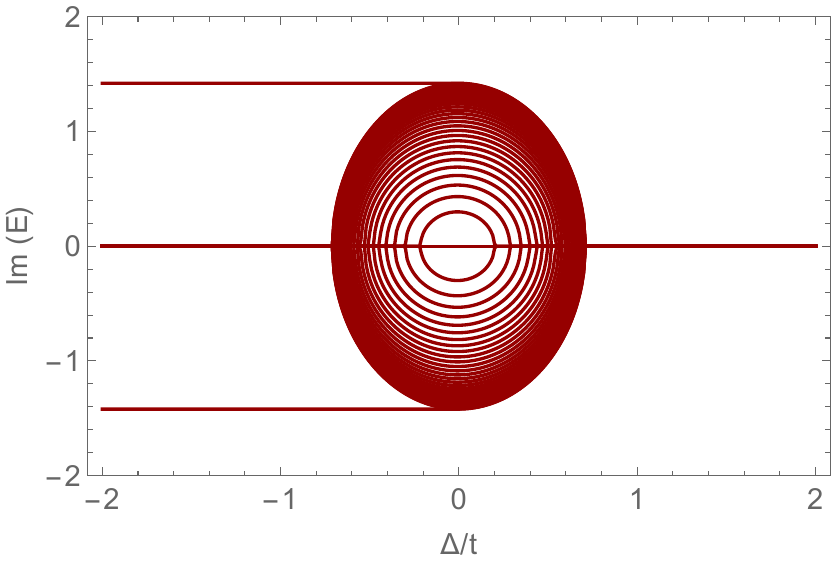}}
   \put(45,35){(c)}
   \end{picture}\\
    \vskip -.4 in
 \begin{picture}(100,100)
     \put(-70,0){
   \includegraphics[width=.45\linewidth]{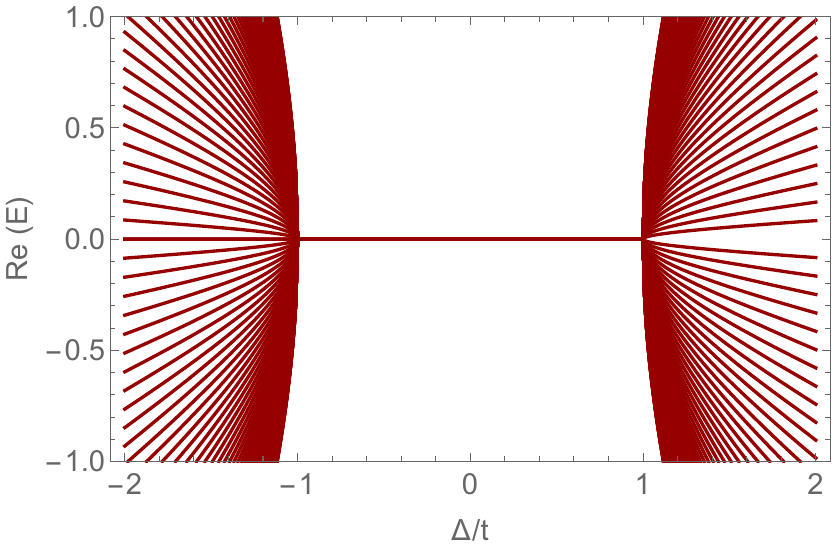}\hskip .2 in
   \includegraphics[width=.45\linewidth]{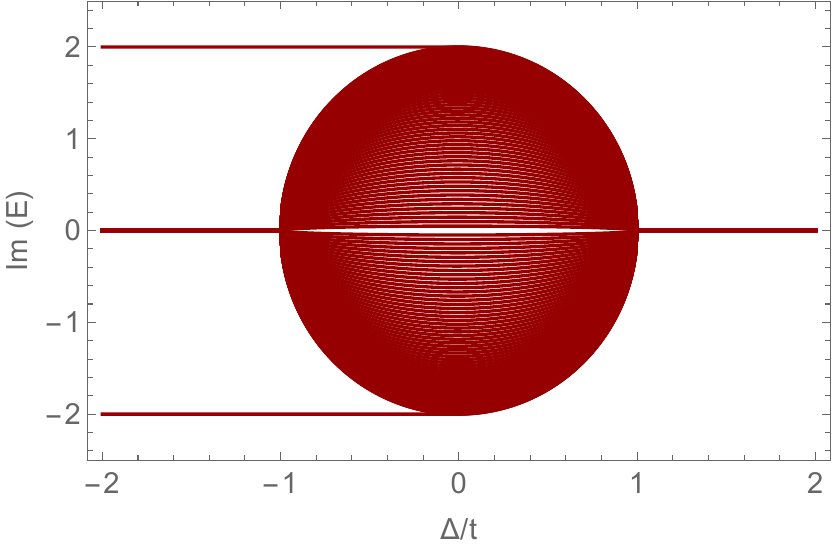}}
   \put(45,35){(d)}
   \end{picture}\\
    \vskip -.4 in  \begin{picture}(100,100)
     \put(-70,0){
   \includegraphics[width=.45\linewidth]{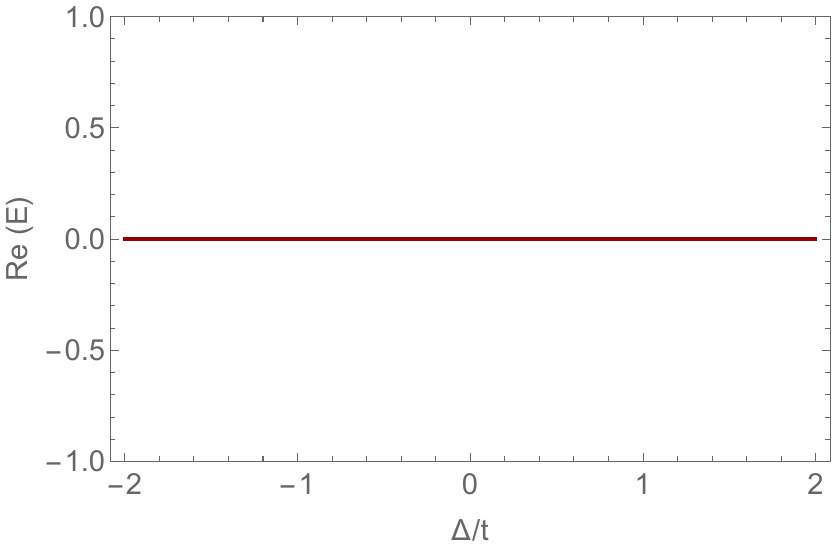}\hskip .2 in
   \includegraphics[width=.45\linewidth]{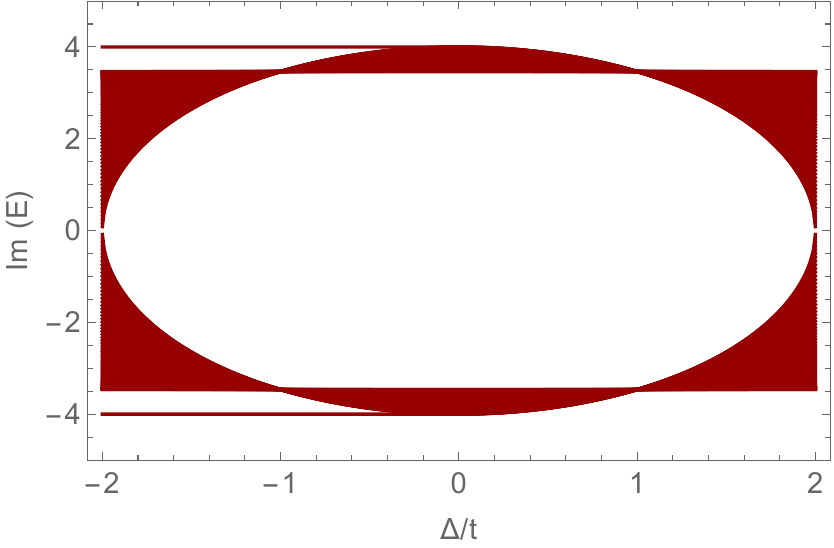}}
   \put(45,35){(e)}
   \end{picture} 
  %\vskip -0.1 in
\caption{Numerical spectra [the {\bf real}(left) and the {\bf imaginary} (right) part of the eigenvalue] of an {\bf open chain} for the $\mathcal{PT}$-symmetric NH SSH model with respect to $\Delta/t$ with $L=128$ and for $\theta=\pi$ and $\gamma/t$~=~ (a) $2.45\times10^{-2}$,(b) $7.32\times10^{-2}$, (c) 1.42, (d) 2 and (e) 4 respectively.} 
\label{fig1}
\end{figure}
in which $k$ is the single-particle momentum. Notice that the Hamiltonian Eq.(\ref{5a}) is $\mathcal{PT}$ symmetric i.e., $(\mathcal{PT})\mathcal{H}^{\star}_{\mathcal{PT}}(k)(\mathcal{PT})^{-1}=\mathcal{H}_{\mathcal{PT}}(k)$ with $\sigma_{x}\mathcal{K}$ being the $\mathcal{PT}$ symmetry operator. The normalized right eigenstates are obtained as 
$$\ket {\psi_{+}}=\begin{pmatrix}
 e^{-i\phi_{k}}\cos\theta_{k} \\
  \sin\theta_{k}
\end{pmatrix}\;,\;\ket {\psi_{-}}=\begin{pmatrix}
 -e^{-i\phi_{k}}\sin\theta_{k} \\
  \cos\theta_{k}
\end{pmatrix}$$
while the normalized left eigenstates are 
$$\ket {\lambda_{+}}=\begin{pmatrix}
 e^{i\phi_{k}}\cos\theta_{k} \\
  \sin\theta_{k}
\end{pmatrix}\;,\;\ket {\lambda_{-}}=\begin{pmatrix}
 -e^{i\phi_{k}}\sin\theta_{k} \\
  \cos\theta_{k}
\end{pmatrix}$$
satisfying $\mathcal{H}_{\mathcal{PT}}(k)\ket{\psi_{\pm}}=\pm E(k)\ket{\psi_{\pm}}$ and  $\mathcal{H}^{\dagger}_{\mathcal{PT}}(k)\ket{\lambda_{\pm}}=\pm E(k)^{\star}\ket{\lambda_{\pm}}$. In these expressions $\phi_{k}=i\ln\Big|\frac{(t+\Delta)+(t-\Delta)e^{-ik}}{|(t+\Delta)+(t-\Delta)e^{-ik}|}\Big|$ and $\theta_{k}=\tan^{-1}\Big(\sqrt{\frac{E_{+}-i\gamma}{E_{+}+i\gamma}}\Big)$. For the NH systems, left and right eigenstates follow the bi-orthogonal condition: $\braket{\lambda_{\pm}|\psi_{\pm}}=1$ and  $\braket{\lambda_{\mp}|\psi_{\pm}}=0$.

\begin{figure}
  \vskip -.1 in
  \begin{picture}(100,100)
     \put(-100,0){
       \includegraphics[width=.55\linewidth,height=1.2 in]{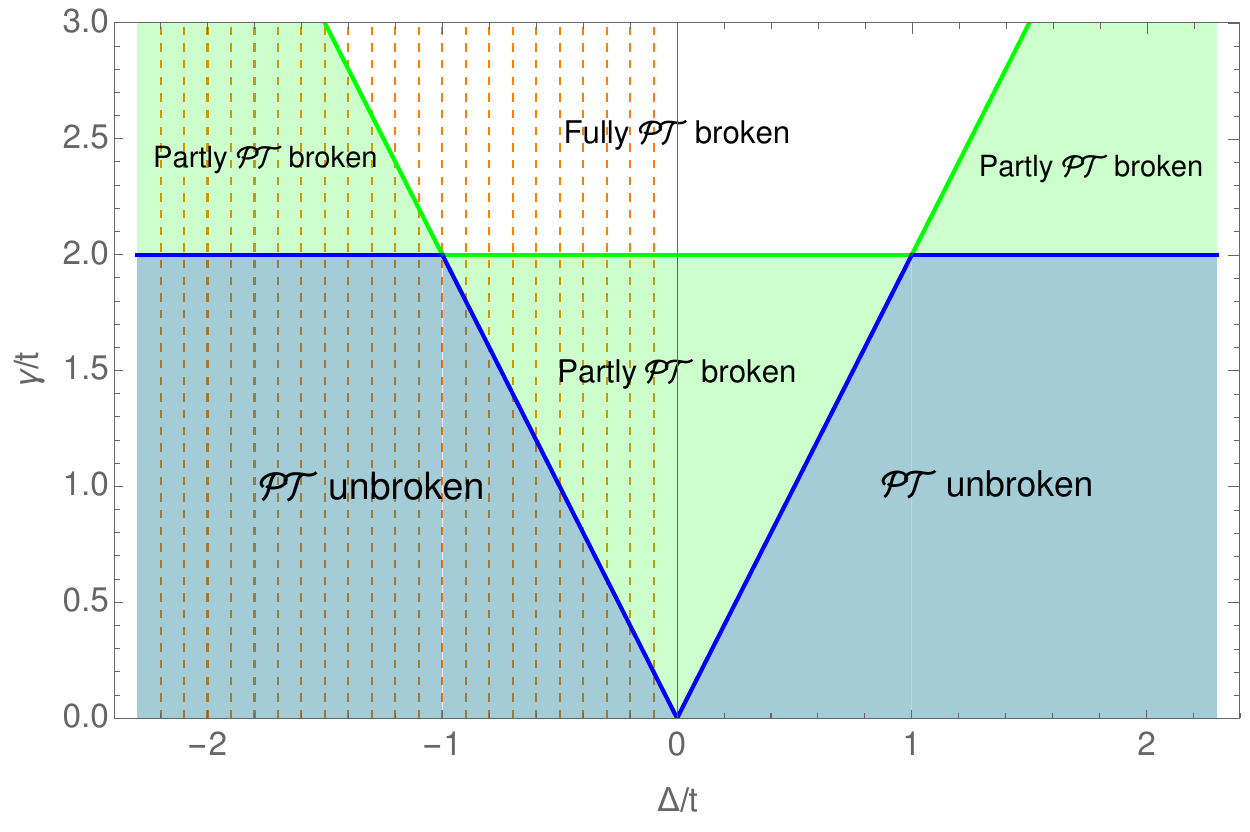} \includegraphics[width=.55\linewidth,height=1.2 in]{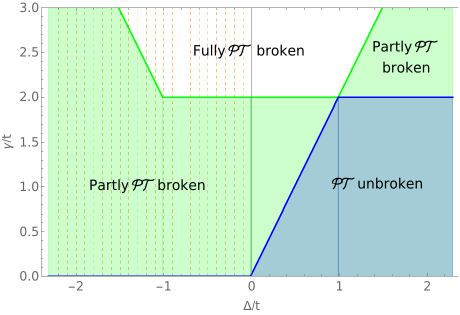}}
     \put(-72,75){(a)}
     \put(56,75){(b)}
  \end{picture}
  \caption{{Phase Diagrams of a NH SSH chain with (a) periodic and (b) open boundary conditions for $\theta=\pi$. All of $\mathcal{PT}$ unbroken, partially broken and fully broken phases are shown in this $\gamma-\Delta$ plot. Also the topological phases are marked as the vertically striped regime.}}
  \label{figph}
  \end{figure}

For a periodic system, a larger $\gamma$ results in a narrower band gap for fixed $\Delta/t$. For {$|\Delta/t|<1$}, all eigenvalues become real for all the Bloch states as long as $\gamma<2|\Delta|$ (and for $\gamma<2t$ when $|\Delta/t|>1$) making the system $\mathcal{PT}$ symmetric\cite{weimann,lang} (see Fig.\ref{fig1}(d), for example) while for $\gamma>2t$ {when $|\Delta/t|<1$} (and for $\gamma>2|\Delta|$ when $|\Delta/t|>1$) one gets imaginary eigenvalues for all $k$ values (see Fig.\ref{fig1}(e)) giving a fully $\mathcal{PT}$-symmetry broken phase. Moreover, the system goes into a partially $\mathcal{PT}$ broken region whenever $2t>\gamma>2|\Delta|$\cite{ep} for $|\Delta/t|<1$ or, when $2t<\gamma<2|\Delta|$ for $|\Delta/t|>1$ {(see the phase diagram in Fig.\ref{figph}(a))} where the energy eigenvalues become real for some values of $k$ and imaginary for the other. For an open (finite) system, however, there are end states in the {topological phase} region and the corresponding eigenvalues go from real to imaginary at a much smaller critical value of $\gamma_{ep}~(<<2|\Delta|)$ called an {\it exceptional point} (EP)\cite{ep} {(see Fig.\ref{figph}(b)). This is different from $\mathcal{SPT~BT}$ occurring in a periodic system when the bulk band gap closes\cite{ep2}}.  The energy spectra under OBC in the topological ($\Delta/t<0$) and trivial phase ($\Delta/t>0$) regimes show features different from the conventional SSH chain in the presence of an onsite imaginary potential $U$ and are presented in Fig.\ref{fig1}.

\subsubsection{{Topologically Nontrivial Phase (TNP)}}

Firstly, we proceed to discuss how the properties of the topological phase get changed with $\gamma$. Notice that the real spectra for small $\gamma=2.45\times10^{-2}$ depicted in Fig.\ref{fig1}(a) has identical behavior to that of the usual SSH model\cite{mandal} and here two (purely)real bulk bands are separated by midgap zero modes which however possess imaginary eigenvalues. For higher $\gamma$, few bulk bands with small real energy start getting their imaginary components gradually (see Fig.\ref{fig1}(c)-(e))\cite{lieu}. Moreover, the mid-gap states (here, we call them NH ZES) having $Re[E]=0$ and conjugated imaginary energies $\pm i\beta$ (where $\beta$ should depend on $\Delta/t$ and $\gamma$) indicate end state behavior\cite{lieu,zhu} throughout the TNP ($i.e.,~\Delta/t<0$). In that sense, they can be called gapped end states. The $\mathcal{PT}$ phase diagram Fig.\ref{fig2}(a) shows the variation of imaginary part of the eigenvalues (within TNP) of the end states with $\gamma$. It indicates appearance of imaginary conjugate pair of eigenvalues beyond $\gamma=\gamma_{ep}$ resulting in a $\mathcal{PT}$ symmetry breaking in the TNP for $\gamma\ge\gamma_{ep}$[Fig.\ref{fig3}(b)]. In contrast, a purely real spectrum is observed in the Kitaev chain as they obey particle-hole symmetry and leads to the $\mathcal{PT}$ symmetric end state\cite{lieu}. This indicates that, as $\gamma$ is introduced and gradually increased in the system, it is the symmetry of the end states that provides the overall $\mathcal{PT}$-symmetry rather than the mere existence of TNP, as also elaborated in Ref.\cite{pt}.
The broken $\mathcal{PT}$-symmetry of the end modes for small $\gamma~(>\gamma_{ep}$ though) is again responsible for destroying the stability of zero-modes of a periodically modulated SSH model in the TNP\cite{zhu}. %One can notice the gapless bulk modes along with gapped end modes.
For the particular value of $\gamma$ as mentioned above, there are only two conjugated eigenvalues corresponding to a gapped end state (as discussed above) in the imaginary spectrum and the remaining $2N-2$ eigenvalues are purely real. Notice that, unlike our chiral symmetry broken system, the end modes in the models respecting chiral symmetry are gapless with zero energies\cite{lieu}. Then for $\gamma=7.32\times10^{-2}$ in Fig.\ref{fig1}(b), few of the gapped bulk modes also get nonzero imaginary component in energy for $\gamma$ close to the topological transition point (compare with the range of $2|\Delta|<\gamma$ for the loss of $\mathcal{PT}$ symmetry in a periodic chain), similar to another $\mathcal{PT}$-symmetric model studied in Ref.\cite{lieu}. Notice that such transition for very small $\gamma$ is not discernible well as it falls very close to the TQPT. The gap between the end state pairs entirely depends on $\gamma$ as $E_{edge}=~\pm i\gamma$ with the gap increasing linearly with $\gamma$ [see right panels of Fig.\ref{fig1}]. Even for the bulk modes whose energy becomes complex (beginning from the situation like in Fig.\ref{fig1}(b)),  the energy gap in the imaginary plane also increases with $\gamma$. {Notice that, in compatible with Eq.\ref{6} and Fig.\ref{figph}(a),} all the bulk energy modes become purely imaginary for {$|\Delta|<\Delta_0=\gamma/2$, like shown in Fig.\ref{fig1}(d),(e) (but not so for Fig.\ref{fig1}(b),(c) where $\gamma/t<1$ and a partially $\mathcal{PT}$ broken phase is obtained). A pair of NH~ZES exist for $\Delta/t<0$ which turn into true ZES at the exceptional point\cite{edge2}. Out of the two,} one end state appears at the left boundary while the other at the right boundary [see Fig.\ref{fig2}(c)]. The properties of NH end states in summarized in Table.\ref{table:2}.

\subsubsection{{Topologically Trivial Phase (TTP)}}
\begin{figure}
   \vskip -.4 in
   \begin{picture}(100,100)
     \put(-70,0){
  \includegraphics[width=.48\linewidth]{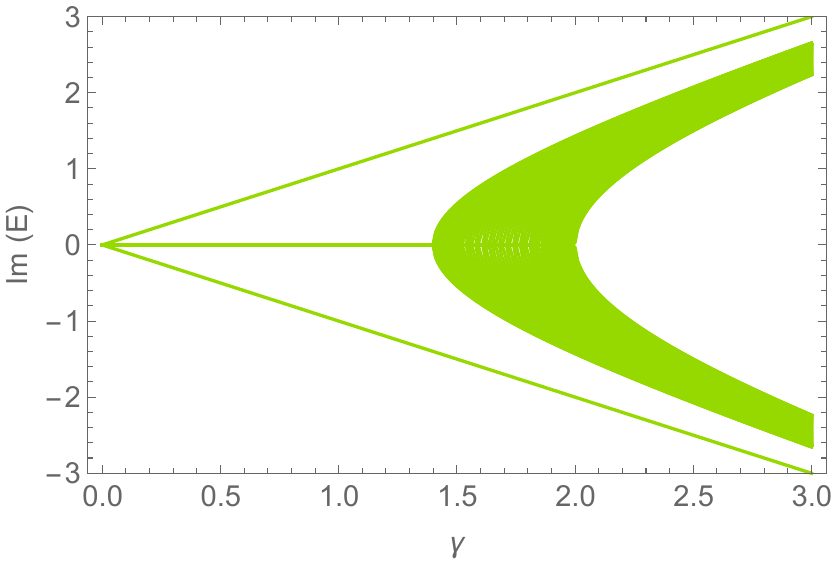}
  \includegraphics[width=.48\linewidth]{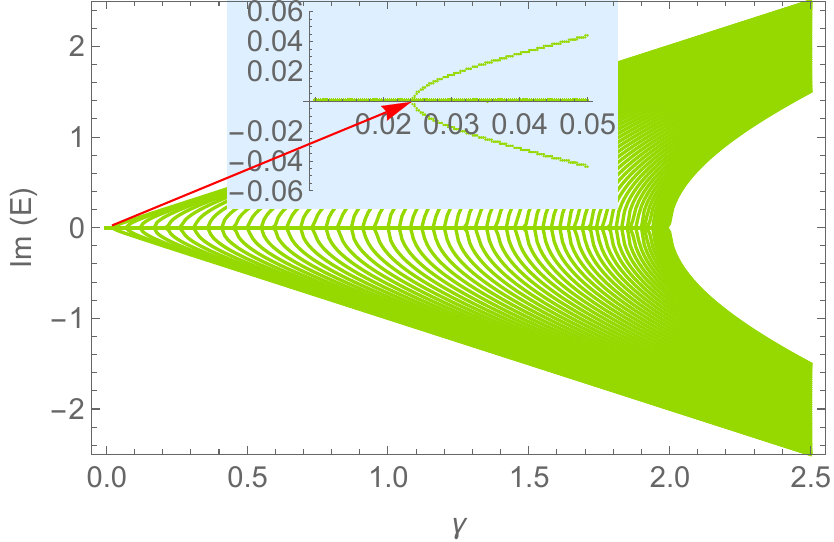}}
     \put(-50,65){(a)}
     \put(75,65){(b)}
   \end{picture}\\
   \vskip -.3 in
   \begin{picture}(100,100)
     \put(-70,0){
   \includegraphics[width=.48\linewidth]{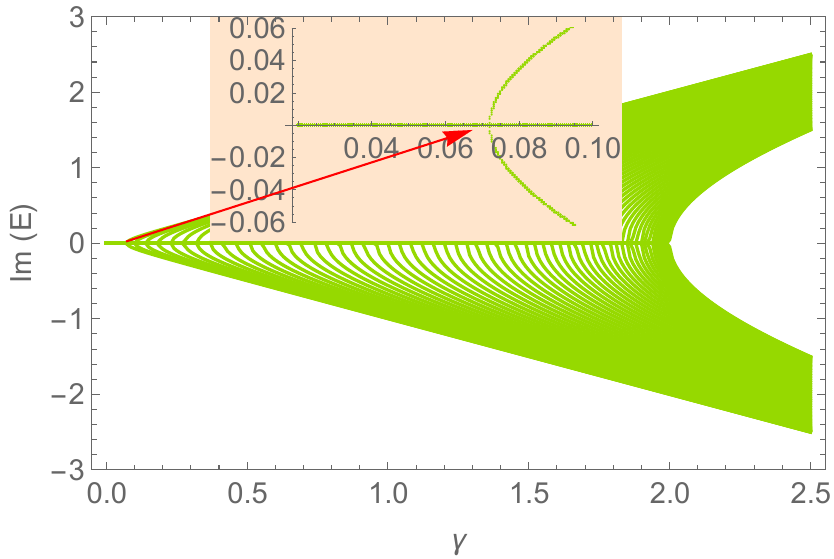}
   \includegraphics[width=.48\linewidth]{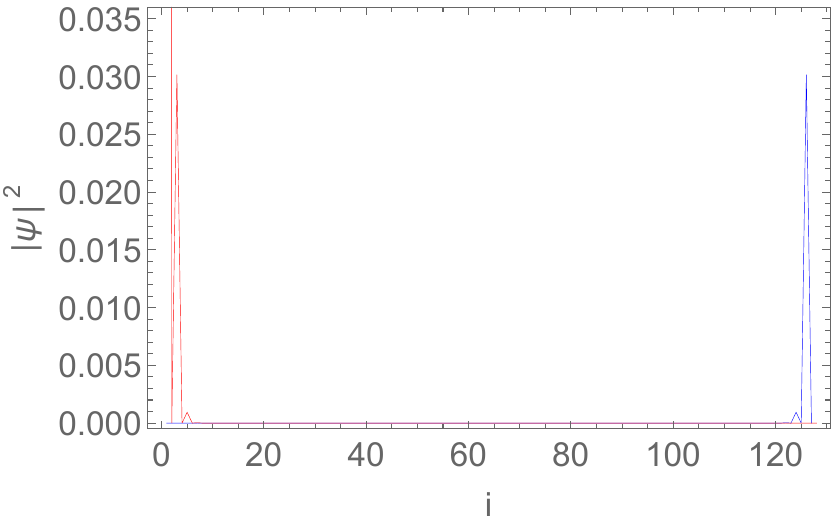}}
   \put(-50,65){(c)}
   \put(80,65){(d)}
    \end{picture} 
  %\vskip -0.1 in
\caption{Numerical spectra of the {\bf imaginary part} of the eigenvalue against $\gamma$ with $t=1$,~$L=128$ and $\Delta/t$=~(a)~$-7\times10^{-1}$ (within TNP), (b)~0 (at TQPT) and (c)~$3.01\times~10^{-2}$ (within TTP). (d) The amplitude distribution of two imaginary end states with $\Delta/t$=$-7.32\times10^{-1}$ (deep within the TNP) and $\gamma>\gamma_{ep}$. Inset illustrates the exact location of $\mathcal{SPT~BT}$ point.} 
\label{fig2}
\end{figure}

Now, we study the {trivial phase} which appears for $\Delta/t>0$. One can notice only gapped bulk modes with entirely real energy eigenvalues for $\gamma=~2.45\times10^{-2}$ [see Fig.\ref{fig1}(a)], retaining the $\mathcal{PT}$-symmetry there. However, the TTP in contrast to the TNP doesn't support any end mode. But other than that, the spectra is symmetric with respect to the sign of $\Delta/t$, even in the presence of $\gamma$. Even though the TTP preserves $\mathcal{PT}$-symmetry for small $\gamma$, they lose the same for larger values of it [see Fig.\ref{fig2}(c)].
Moreover, Fig.\ref{fig2}(b) depicts that the imaginary eigenvalues start to appear at the phase boundary of $\Delta/t\rightarrow~0$\cite{comment1} at $\gamma=~2.45\times~10^{-2}$. 

\begin{figure}
   \vskip -.4 in
   \begin{picture}(100,100)
     \put(-70,0){
     \includegraphics[width=.48\linewidth]{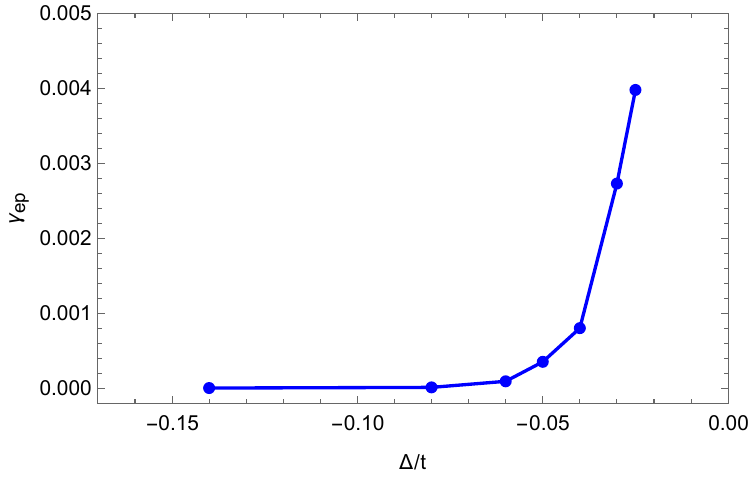}
  \includegraphics[width=.48\linewidth]{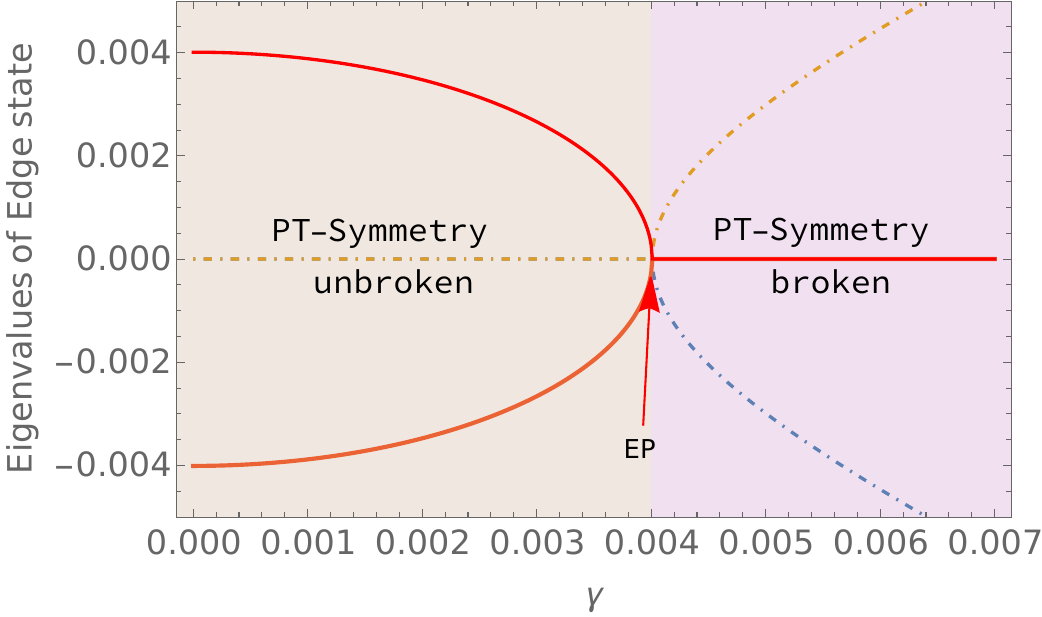}}
     \put(-45,52){(a)}
     \put(80,52){(b)}
   \end{picture}\\
   \vskip -.2 in
   \begin{picture}(100,100)
     \put(-70,0){
     \includegraphics[width=.48\linewidth]{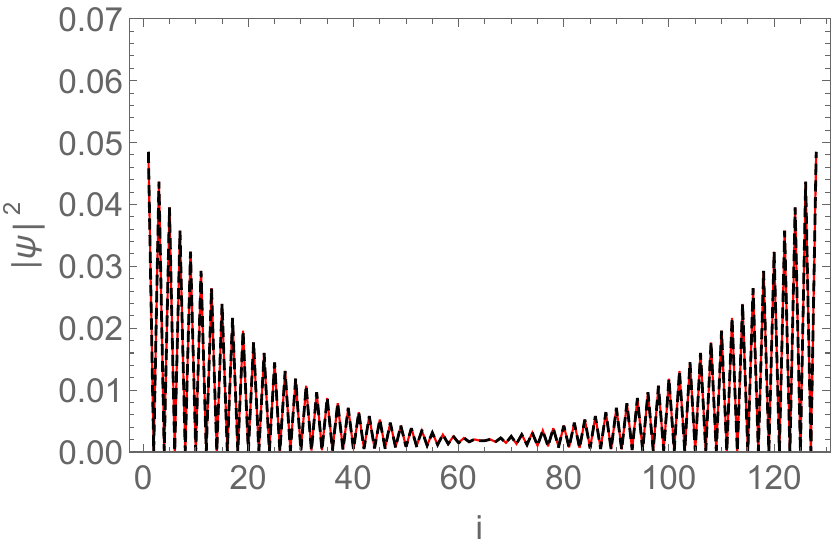}
   \includegraphics[width=.48\linewidth]{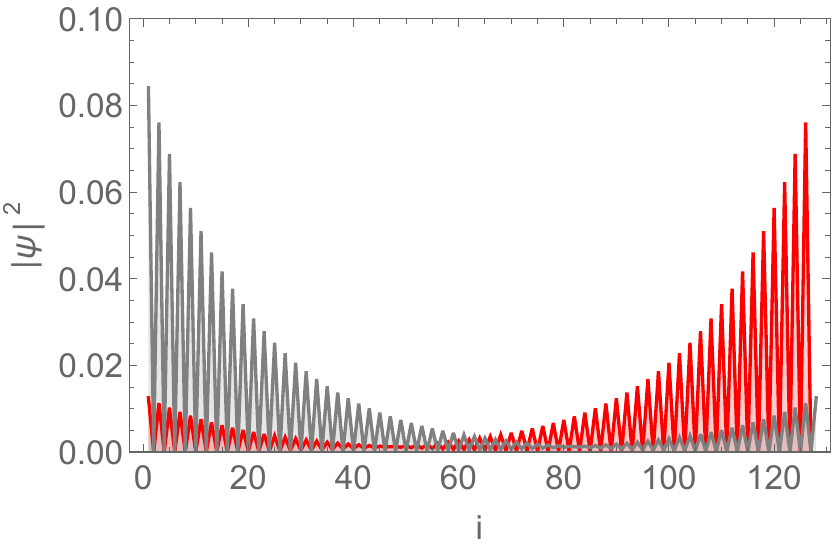}}
   \put(-45,52){(c)}
    \put(80,52){(d)}
    \end{picture} 
  %\vskip -0.1 in
\caption{(a)Variation of $\gamma_{ep}$ with $\Delta$ in an open system. (b) {Coalescence of end state eigenvalues within} TNP at $\gamma_{ep}$ for $\theta=\pi$. The solid and dashed lines show the real and imaginary components of energy respectively. The behavior of end states at lattice site $i$ (absolute values of amplitudes) for smaller (c) and greater (d) than $\gamma_{ep}$. The chosen parameters are ~$L=128$ and $\Delta/t$~=~$-2.5\times10^{-2}$ (very close to the TQPT).} 
\label{fig3}
\end{figure}

As we studied above, the nonzero $\gamma$ values results in a $\mathcal{SPT~BP}$ within the topological phase as shown in Fig.\ref{fig2}, though followed by a $\mathcal{SPT~BT}$ at a small value of $\gamma=\gamma_{ep}$ at which the end state eigenvalues collapse to zero\cite{ep}. $\gamma_{ep}$ tends to zero in the infinite chain limit\cite{ep}. {In a finite chain instead, a small value for $\gamma_{ep}$ can be found close to TQPT which practically goes to zero as $\Delta/t$ is varied towards deep within TNP.} Fig.\ref{fig3}(a) illustrates how $\gamma_{ep}\sim 0$ for $\Delta/t<<0$. {Hence in the phase diagram for a chain with OBC, $\mathcal{PT}$ symmetry is broken within TNP practically immediately after switching on $\gamma$ (Fig.\ref{figph}(b)).} We also demonstrated the behavior of the end state eigenvalues across EP at a point $\Delta/t=-2.5\times 10^{-2}$ which is within TNP and very close to the TQPT point\cite{comment} (see Fig.\ref{fig3}(b)). Here, the system shows entirely real (imaginary) end-state spectra leading to $\mathcal{PT}$ unbroken (broken) phase for $\gamma~<(>)\gamma_{ep}$. {Numerical diagonalization of Eq.(\ref{1}) shows that value to be} $\gamma_{ep}/t=~3.98\times10^{-3}$.

The above value of $\mathcal{PT}$-symmetry breaking point, $\gamma_{ep}$ can be verified with analytical treatments {probing the coupling between the end states as} proposed in Ref.\cite{ep}. The plot of end states displays that they have an equal share on both ends for $\gamma<\gamma_{ep}$ [Fig.\ref{fig3}(c)] while for $\gamma>\gamma_{ep}$, NH end states emerge\cite {edge1,edge2,end,end1} with imaginary eigenvalues. These NH end states tilt towards one end [see Fig.\ref{fig3}(d)], {unlike their Hermitian counterpart}. The bulk states remain delocalized or extended and our model not being non-reciprocal, the NH skin effect\cite{zhang,yan} is not found here {(though few bulk modes in the maximally dimerized limit and the in-gap modes show localized states at edges, as we show later in this paper)}. We can, however, add here that under strong dimerization ($i.e.,~\Delta\sim t$), this NH model can exhibit simple harmonic dynamics at EP\cite{sho}.

\subsubsection{Winding Number and Zak Phase}
It is worthwhile to mention here that one can estimate the winding number (considering the same form as in the Hermitian case\cite{edge2}) for the periodic system when the bulk band gap appears\cite{Schomerus}. One can estimate the winding number from the following formula\cite{edge2,mandal}
\begin{equation}\label{winding}
\mathcal{W}=\frac{1}{2\pi i}\int_{BZ}q(k)^{-1}\partial_{k}q(k)dk
\end{equation}
where the complex function $q(k)$ is the upper off-diagonal block of the matrix $Q(k)$\cite{edge2,comment2} and defined as
\begin{equation}\label{winding2}
Q(k)=
\begin{pmatrix}
0 & q(k) \\
 q(k)^{\star} & 0
\end{pmatrix},
\end{equation}
However, $Q(k)$ is defined from the NH generalizations of projection operators via $Q(k)=Q^{\prime}(k)+Q^{\prime}(k)^{\dagger}$ with $Q^{\prime}(k)=\frac{1}{2}\Big(\sum_{n>0}\ket{\lambda_{n}(k)}\bra{\psi_{n}(k)}-\sum_{n<0}\ket{\lambda_{n}(k)}\bra{\psi_{n}(k)}\Big)$ in which $n=\pm$ represents the different band index, $\ket{\lambda_{n}(k)}$ and $\ket{\psi_{n}(k)}$ are the left and right eigenstates, respectively, of $\mathcal{H}_{\mathcal{PT}}(k)$. Using Cauchy residue theorem, we can arrive at $\mathcal{W}=1(0)$ for $\Delta/t<0(>0)$ regardless of the values of $\gamma$\cite{ghatak}. As per bulk-energy correspondence, the nonzero winding number predicts ZES (in the real part of energy eigenvalues) on the boundary\cite{palyi}, and in our system, this occurs for $\Delta/t<0$. More precisely, the end state with energy $Re[E]=0$ is estimated in the region of $\Delta/t<0$ provided that $\gamma<2|\Delta|$ such that the bulk energy gap remains open\cite{ep}.

Like other dissipative cases\cite{operator}, the Zak phase for our dissipative systems can also be generalized. For unbroken $\mathcal{PT}$ symmetric cases (with all eigenvalues of $\mathcal{H}_{\mathcal{PT}}$ real), the complex Zak phase corresponding to the energies $E_{\pm}$ can be estimated by the bi-orthogonal eigenstates of $\mathcal{H}_{\mathcal{PT}}$, $\bra{\lambda_{\pm}}$ and $\ket{\psi_{\pm}}$, as \cite{zak1,zak2}
\begin{equation}\label{zak}
{Z}^c_{\pm}=i\int_{BZ}\bra{\lambda^{\pm}_{k}}\partial_{k}\ket{\psi^{\pm}_{k}}dk,
\end{equation}
which gives the value of the complex Zak phase of {$\pi$} for $\Delta/t<0$ and 0 for $\Delta/t>0$, irrespective of $\gamma$. Thus, the complex zak phase depicts similar behavior to that of $\mathcal{W}$, although the numerical value has to be differed by a factor of {$\pi$}\cite{lieu,sudin}.

For our concerned $\mathcal{PT}$ symmetric model, the reciprocity of the hoppings are maintained despite its non-Hermiticity. Hence the conventional bulk-boundary condition remains intact resulting in absence of any NH skin effect\cite{sudin}.

\subsection{Case of $\theta=~\pi/2$}
Let us now study the properties of nontrivial and trivial phases for this case when non-Hermiticity is induced by switching on the parameter $\gamma$. {The corresponding Bloch Hamiltonian} in the $k$-space becomes
\begin{equation}\label{7}
\mathcal{H}_{\mathcal{PT}}(k) = 
\begin{pmatrix}
i\gamma & (t+\Delta) & 0 & te^{-4ik} \\
(t+\Delta) & -i\gamma & t & 0 \\
0 & t & i\gamma & (t-\Delta) \\
te^{4ik} & 0 & (t-\Delta) & -i\gamma
\end{pmatrix},
\end{equation}
Again in terms of Pauli matrices, Eq.(\ref{7}) can be written as :
%\begin{multline}\label{7a}
%\begin{widetext}
{\small
\begin{align}\label{7a}
 &\mathcal{H}_{\mathcal{PT}}(k)=i\gamma\sigma_{0}\otimes\sigma_{z}+t\sigma_{0}\otimes\sigma_{x}+\Delta\sigma_{z}\otimes\sigma_{x}+\frac{t}{2}(1+\cos4k)\nonumber\\&\sigma_{x}\otimes\sigma_{x}+ \frac{t}{2}(1-\cos4k)\sigma_{y}\otimes\sigma_{y}+\frac{t}{2}\sin4k(\sigma_{x}\otimes\sigma_{y}+\sigma_{y}\otimes\sigma_{x})
 %\end{multline}}
\end{align}}
%\end{widetext}
with energy eigenvalues 
\begin{equation}\label{8}
  \scriptsize
E(k)=\pm\sqrt{2t^2+\Delta^2-\gamma^2\pm t\sqrt{2t^2+6\Delta^2+2(t^2-\Delta^2) \cos4k}}.
\end{equation}
\begin{figure}[htp]
    \vskip -.6 in
   \begin{picture}(100,100)
     \put(-70,0){
  \includegraphics[width=.45\linewidth,height= .75 in]{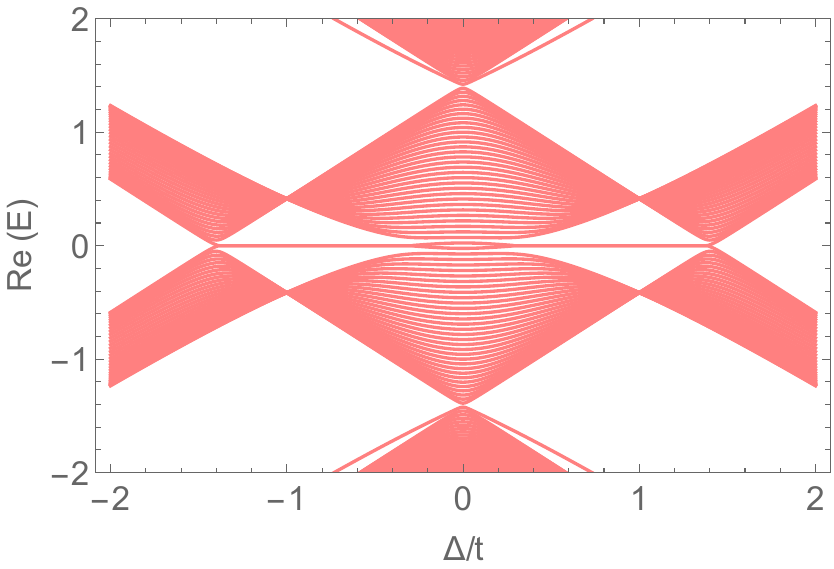}\hskip .2 in
  \includegraphics[width=.45\linewidth,height= .75 in]{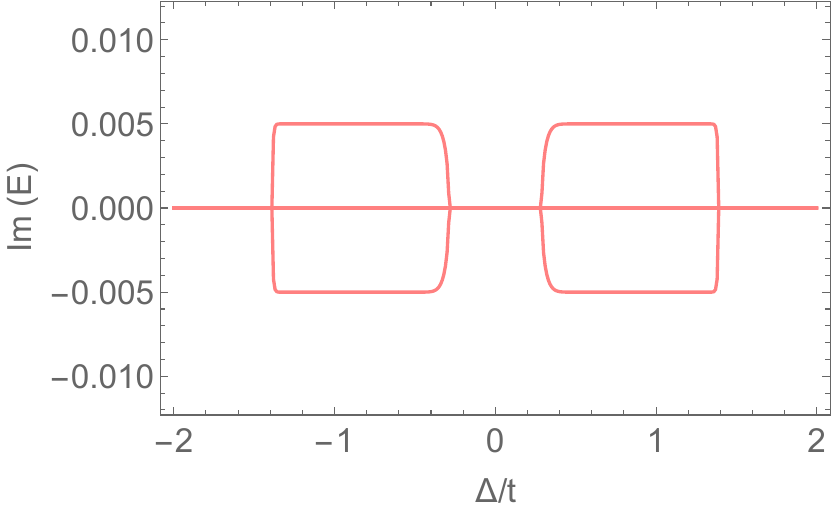}}
     \put(47,30){(a)}
   \end{picture}\\
   \vskip -.6 in
   \begin{picture}(100,100)
     \put(-70,0){
   \includegraphics[width=.45\linewidth,height= .75 in]{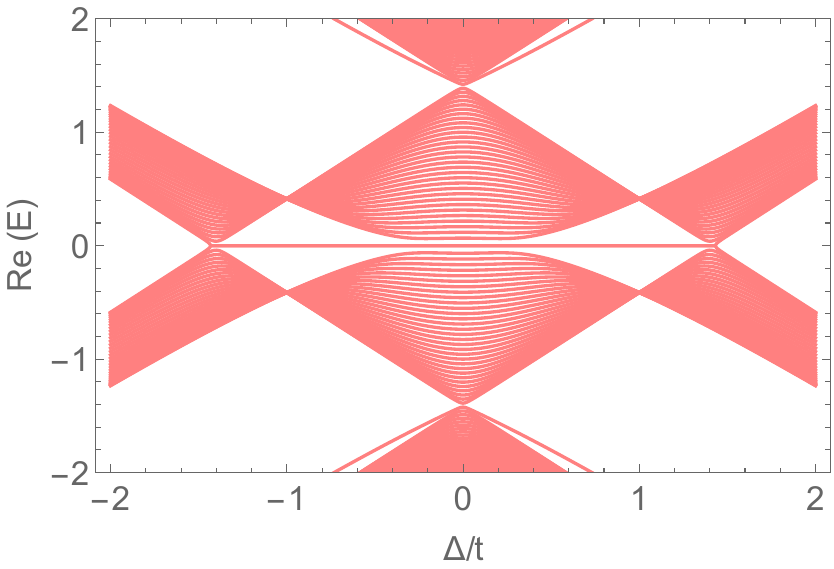}\hskip .2 in
   \includegraphics[width=.45\linewidth,height= .75 in]{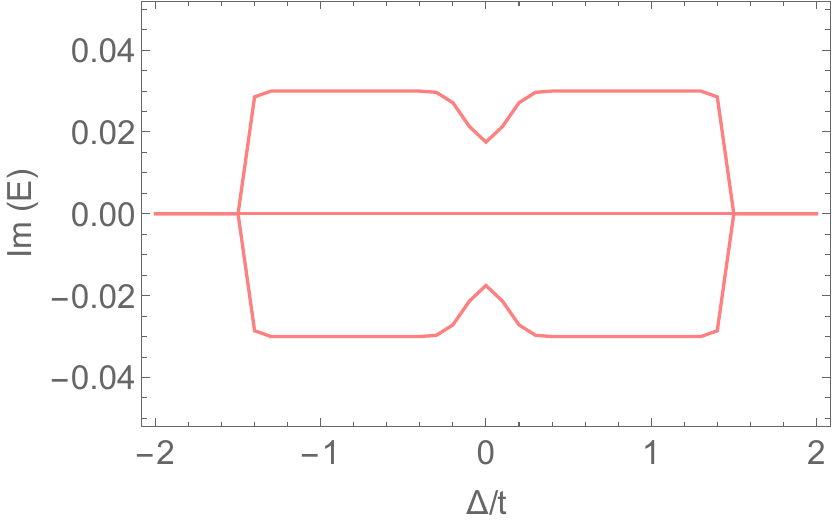}}
   \put(47,30){(b)}
   \end{picture}\\
    \vskip -.6 in
  \begin{picture}(100,100)
     \put(-70,0){
   \includegraphics[width=.45\linewidth,height= .75 in]{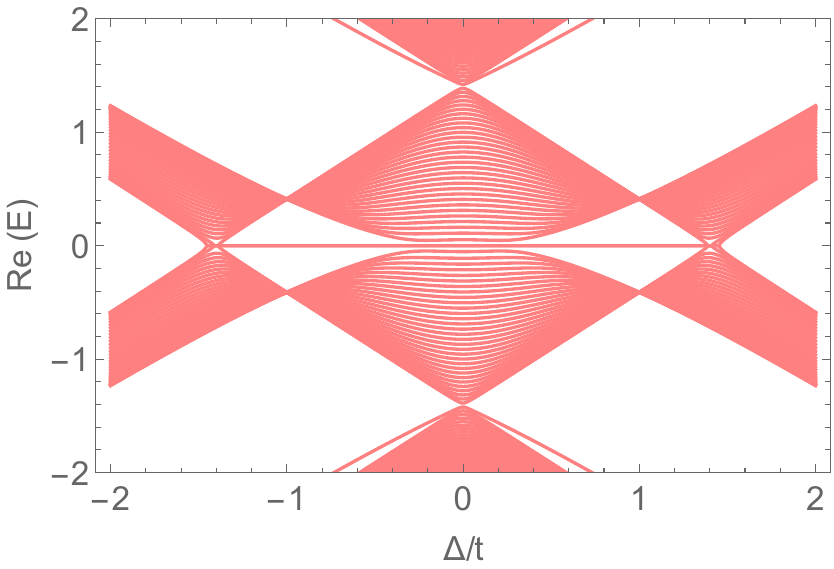}\hskip .2 in
   \includegraphics[width=.45\linewidth,height= .75 in]{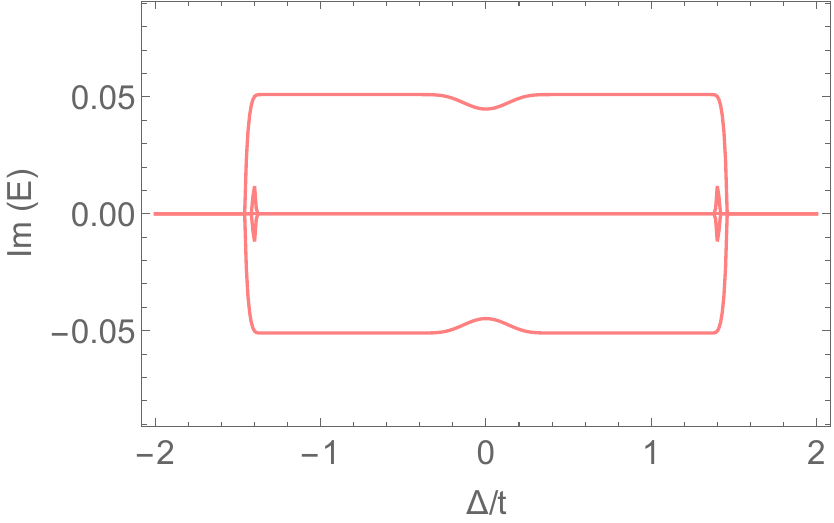}}
   \put(47,30){(c)}
   \end{picture}\\
    \vskip -.6 in \begin{picture}(100,100)
     \put(-70,0){
   \includegraphics[width=.45\linewidth,height= .75 in]{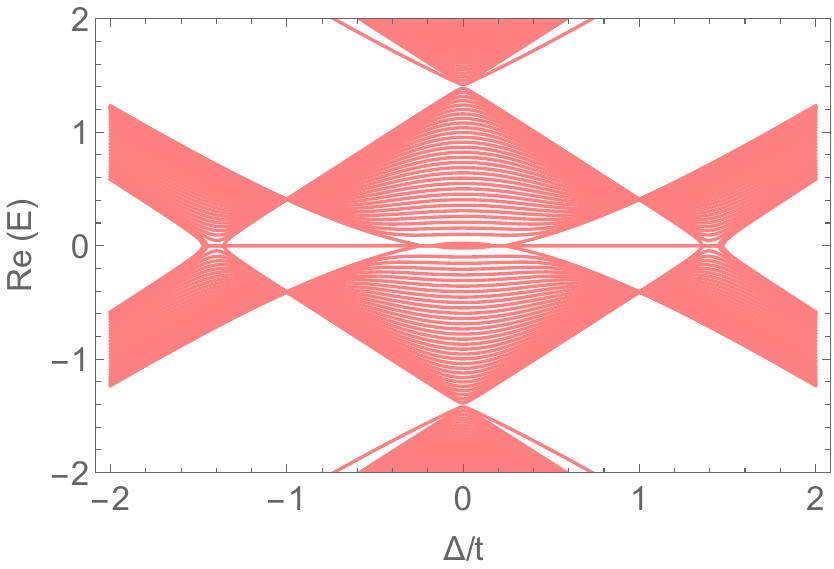}\hskip .2 in
   \includegraphics[width=.45\linewidth,height= .75 in]{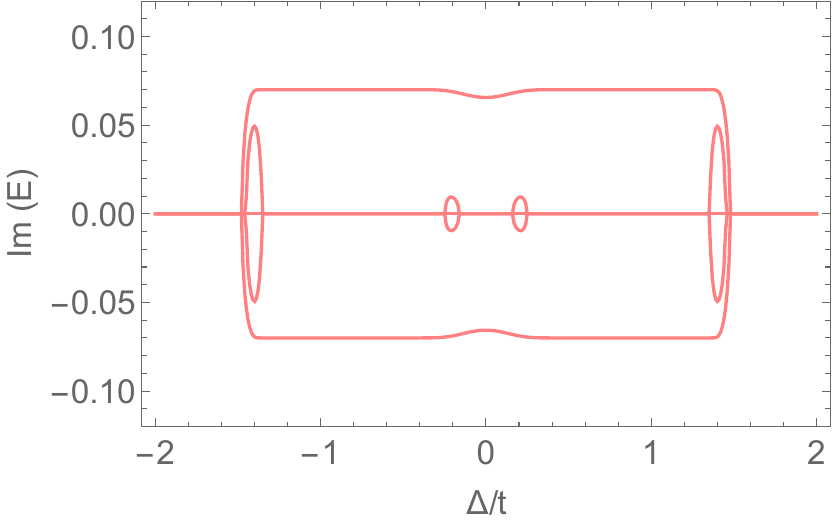}}
   \put(47,30){(d)}
   \end{picture}\\
    \vskip -.6 in
 \begin{picture}(100,100)
     \put(-70,0){
   \includegraphics[width=.45\linewidth,height= .75 in]{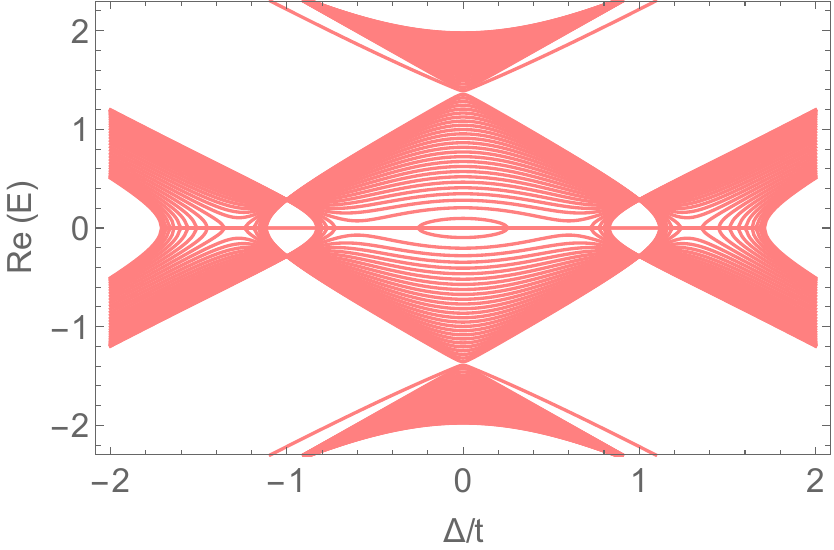}\hskip .2 in
   \includegraphics[width=.45\linewidth,height= .75 in]{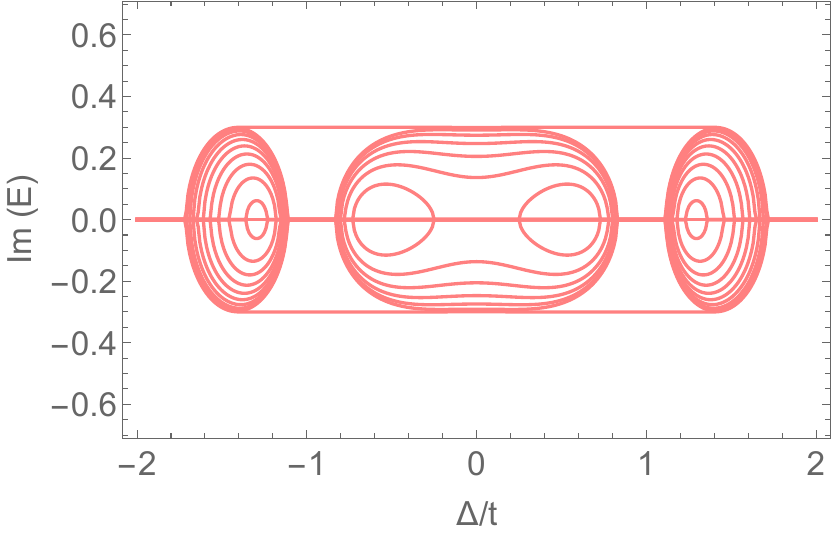}}
   \put(47,30){(e)}
   \end{picture}\\
    \vskip -.6 in  \begin{picture}(100,100)
     \put(-70,0){
   \includegraphics[width=.45\linewidth,height= .75 in]{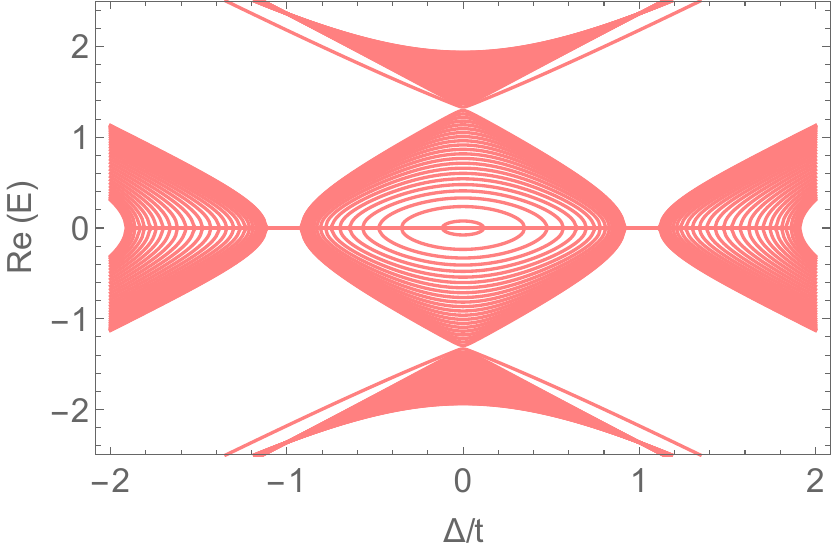}\hskip .2 in
   \includegraphics[width=.45\linewidth,height= .75 in]{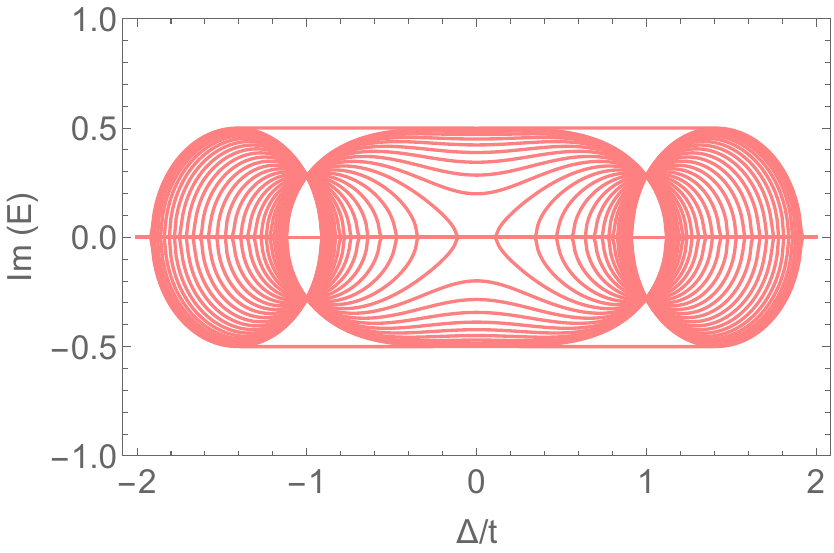}}
   \put(47,30){(f)}
    \end{picture}
   \vskip -.6 in
   \begin{picture}(100,100)
     \put(-70,0){
  \includegraphics[width=.45\linewidth,height= .75 in]{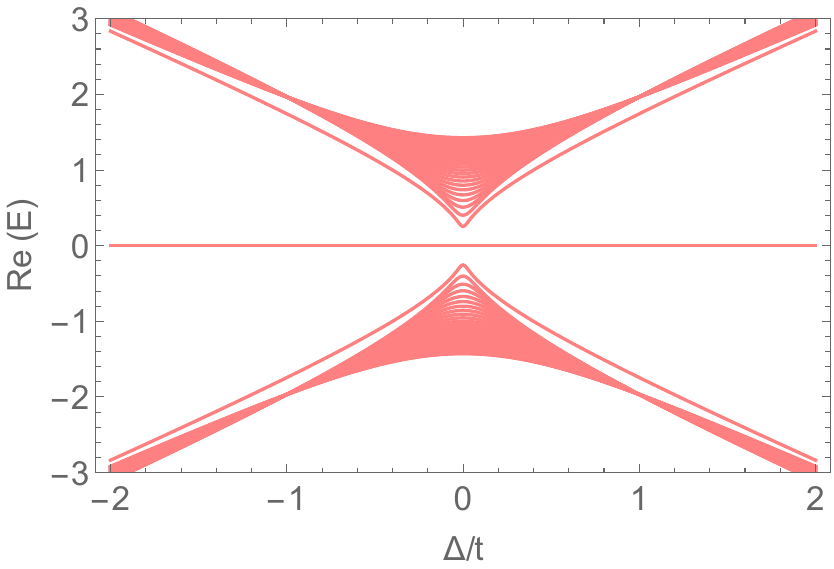}\hskip .2 in
  \includegraphics[width=.45\linewidth,height= .75 in]{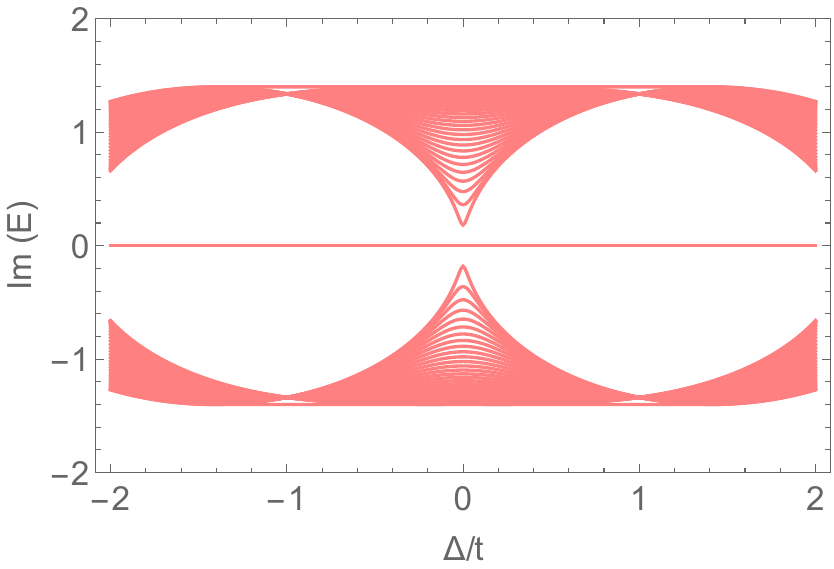}}
     \put(47,30){(g)}
   \end{picture}\\
   \vskip -.6 in
   \begin{picture}(100,100)
     \put(-70,0){
   \includegraphics[width=.45\linewidth,height= .75 in]{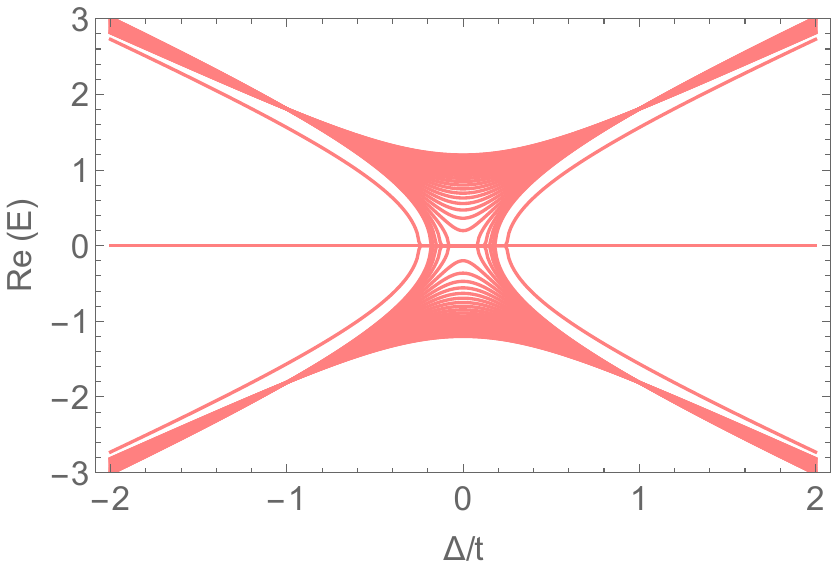}\hskip .2 in
   \includegraphics[width=.45\linewidth,height= .75 in]{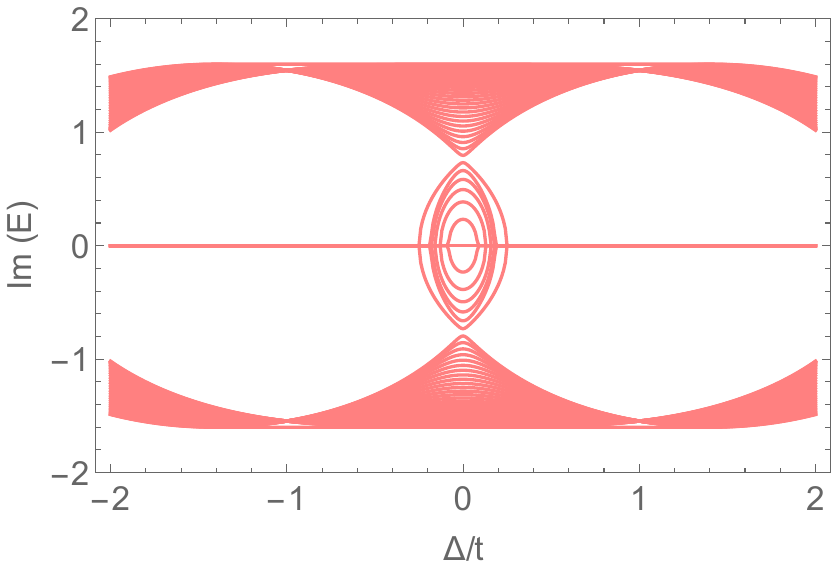}}
   \put(47,30){(h)}
   \end{picture}\\
   \vskip -.6 in
   \begin{picture}(100,100)
     \put(-70,0){
   \includegraphics[width=.45\linewidth,height= .75 in]{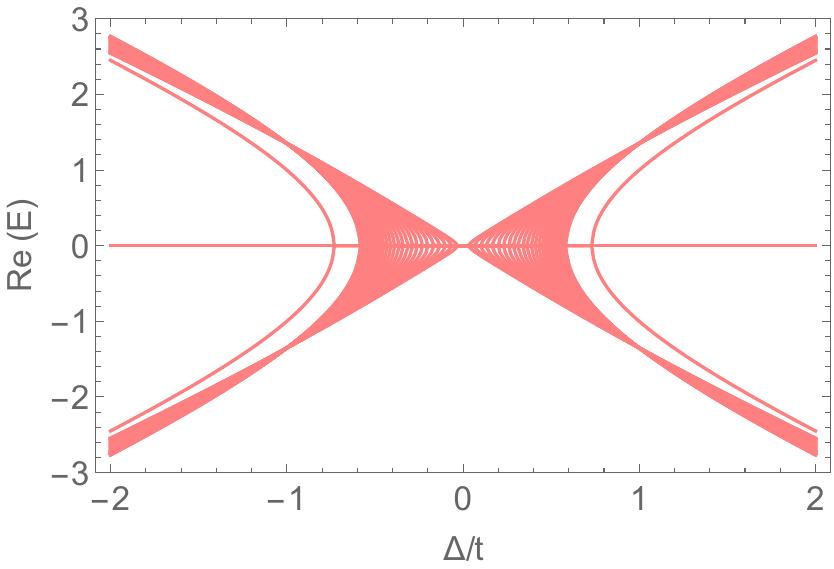}\hskip .2 in
   \includegraphics[width=.45\linewidth,height= .75 in]{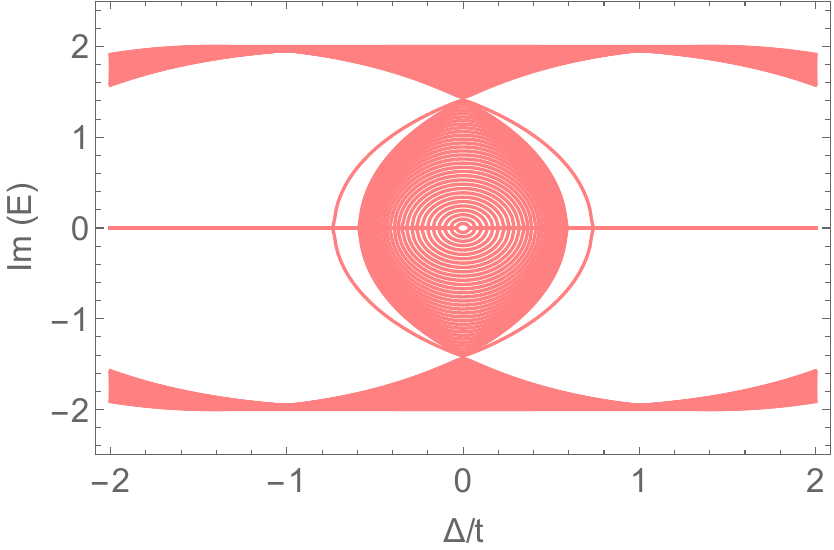}}
   \put(47,30){(i)}
   \end{picture}\\
    \vskip -.6 in
 \begin{picture}(100,100)
     \put(-70,0){
   \includegraphics[width=.45\linewidth,height= .75 in]{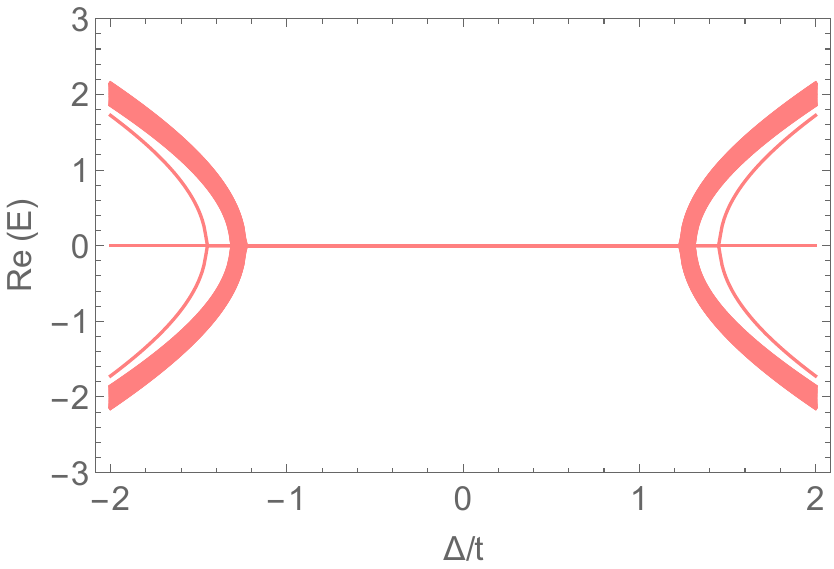}\hskip .2 in
   \includegraphics[width=.45\linewidth,height= .75 in]{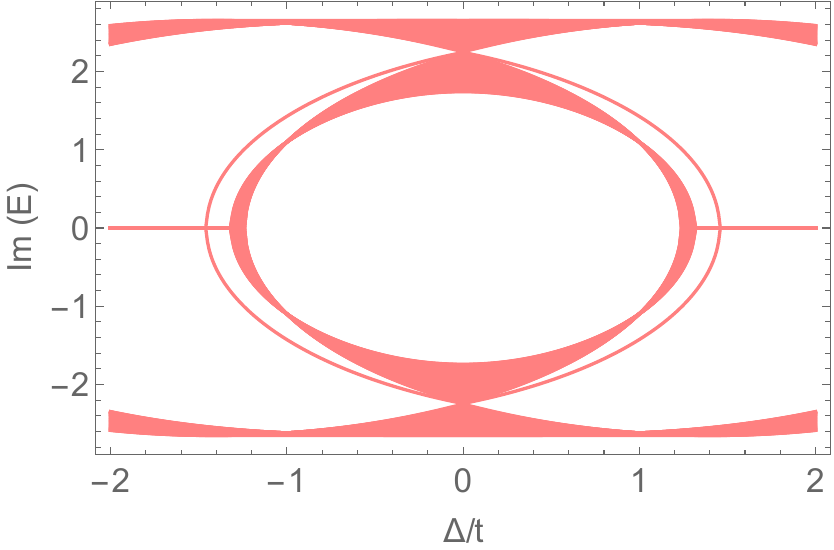}}
   \put(47,30){(j)}
   \end{picture}\\
    \vskip -.6 in  \begin{picture}(100,100)
     \put(-70,0){
   \includegraphics[width=.45\linewidth,height= .75 in]{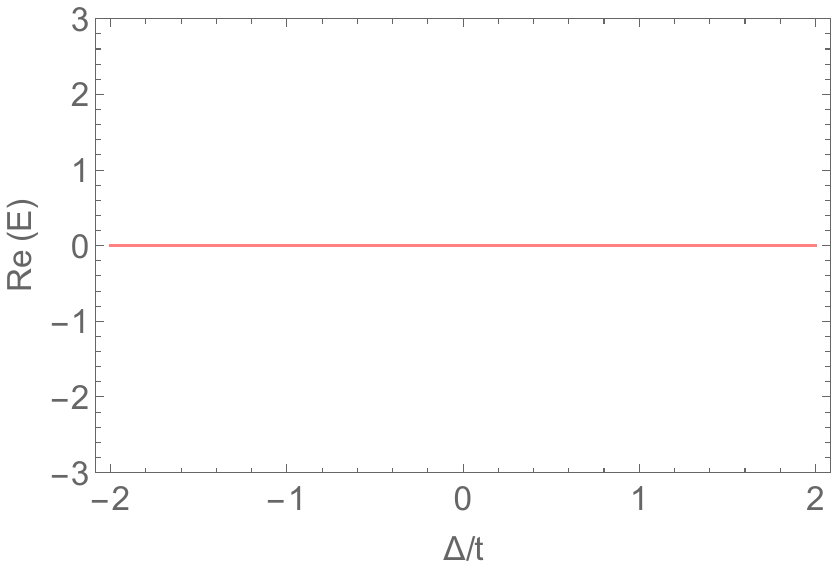}\hskip .2 in
   \includegraphics[width=.45\linewidth,height= .75 in]{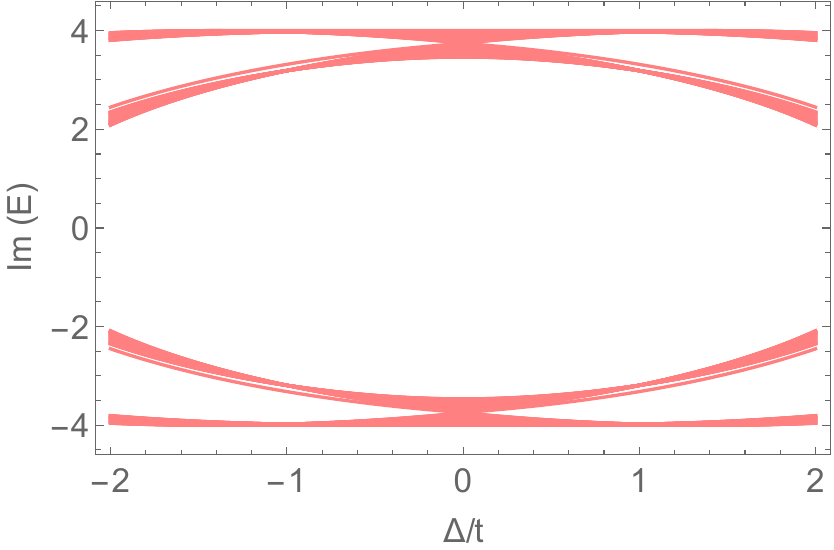}}
   \put(47,30){(k)}
    \end{picture}
  %\vskip -0.1 in
\caption{Numerical spectra of the {\bf real} (left) and {\bf imaginary} (right) part of the energy eigenvalues of an {\bf open chain} for the $\mathcal{PT}$-symmetric NH SSH model with $\Delta/t$ ($\in[-2,~2]$) with $t=1$,~$L=128$ and for various $\gamma$~=~ (a)$5\times10^{-3}$;~(b)$3\times 10^{-2}$;~(c)$5.1\times10^{-2}$;~(d)$7\times10^{-2}$;~(e)$3\times10^{-1}$;~(f)$5\times10^{-1}$;~(g)1.4;~(h)1.6;~(i)2;~(j)2.652;~(k)4.} 
\label{fig4}
\end{figure}
Here, Eq.(\ref{7a}) shows $\mathcal{PT}$ symmetry with $\mathcal{\hat{PT}}=\sigma_{x}\otimes\sigma_{x}\mathcal{K}$. The above dispersion relation gives the $\mathcal{PT}$ classification of the bulk modes.  For $|\Delta/t|<(>)~1$, let's define parameters $\gamma_1$ and $\gamma_2$ as $2t^2+\Delta^2-\gamma^2_{1}=2t\sqrt{t^2+\Delta^2}(2t\sqrt{2\Delta^2})$ and $\gamma^2_{2}=2t^2+\Delta^2+2t\sqrt{t^2+\Delta^2}(2t\sqrt{2\Delta^2})$. Then all the bulk modes become real when $\gamma<|\gamma_1|$ and bulk states show entirely imaginary eigenvalues for $\gamma>|\gamma_2|$. However, the system becomes partly $\mathcal{PT}$ symmetric with energy eigenvalues becoming real for some values of k
  and imaginary for the rest when $|\gamma_{1}|<\gamma<|\gamma_{2}|$ (Fig.\ref{ph-diag2}). We can notice that the gain and loss strength make the efficient modulation of the band gaps for the real and imaginary parts of the energy eigenvalues.

\subsubsection{{Topological Phase}}
\begin{figure}[t]
   \vskip -.4 in
   \begin{picture}(100,100)
     \put(-70,0){
  \includegraphics[width=.48\linewidth]{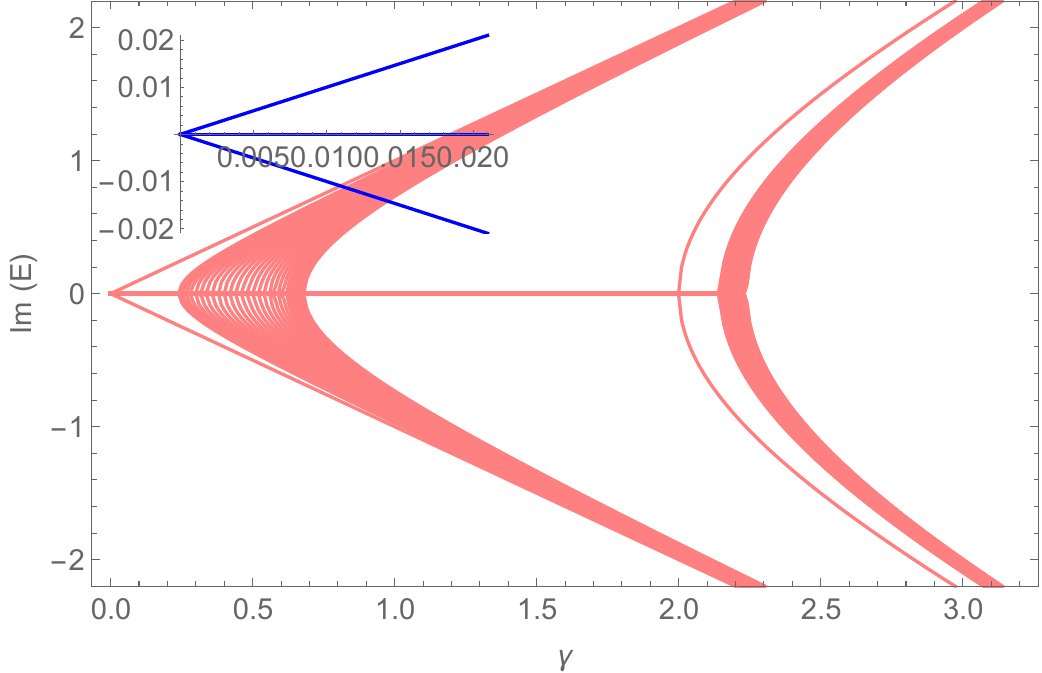}
  \includegraphics[width=.48\linewidth]{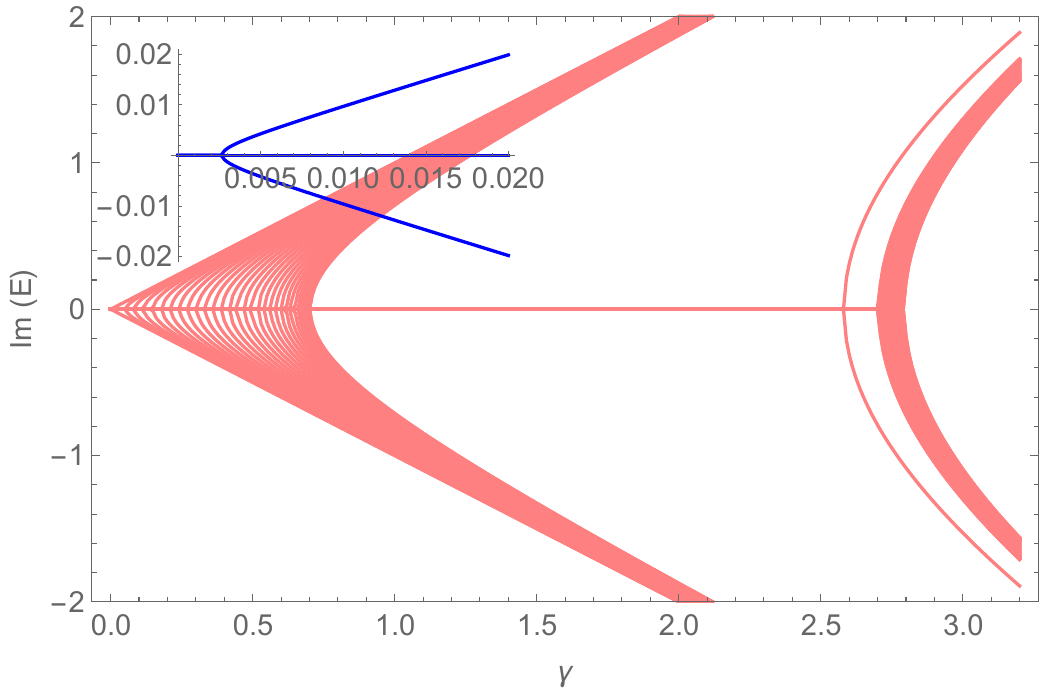}}
     \put(-50,15){(a)}
     \put(80,15){(b)}
   \end{picture}\\
   \vskip -.2 in
   \begin{picture}(100,100)
     \put(-70,0){
   \includegraphics[width=.48\linewidth]{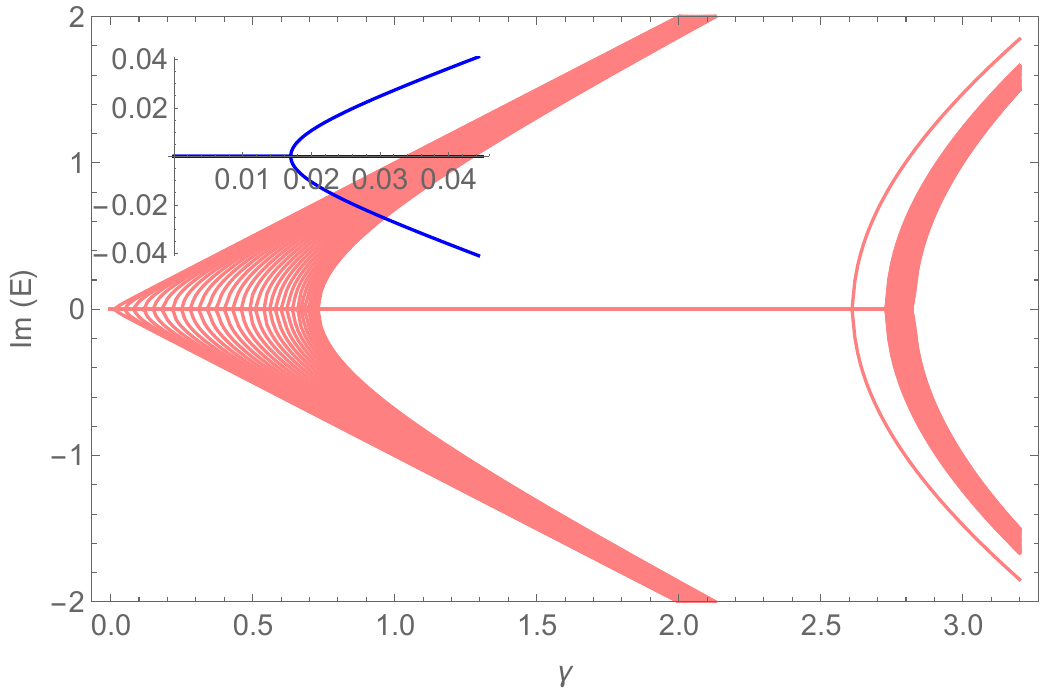}
   \includegraphics[width=.48\linewidth]{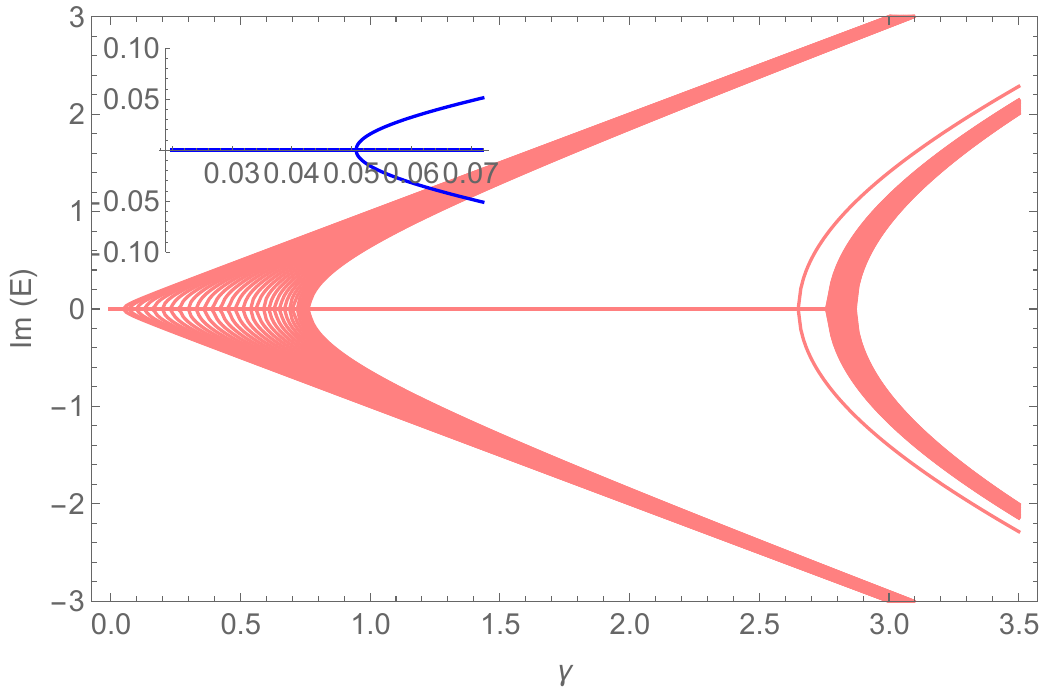}}
   \put(-50,15){(c)}
   \put(80,15){(d)}
    \end{picture} 
  %\vskip -0.1 in
\caption{Numerical spectra of the {\bf imaginary part} of the eigenvalue under {\bf OBC} with respect to $\gamma$ with $t=1$,~$L=128$ and $\Delta/t$~=~(a)~$7.32\times10^{-1}$ (system in the TNP), (b)~1.38 (close to TQPT),~(c)~1.414 (at TQPT) and (d)1.456 (system in the TTP). The inset shows the appropriate position of $\mathcal{SPT~BT}$ point.}
\label{fig5}
\end{figure}

First, we discuss the {topological phase}, i.e., the regime of $0<|\Delta/t|<\sqrt{2}$ (see Table.\ref{table:1}). The variation of the real and imaginary part of energy with $\Delta/t$ under OBC for different values of $\gamma$ is shown in Fig.\ref{fig4}.  Similar to the situation of $\theta=\pi$, here for small $\gamma$ (=~$5\times10^{-3}$) values, the TNP consists of a pair of purely imaginary gapped end states along with purely real bulk modes [Fig.\ref{fig4}(a)]. In order to get a bit clearer vision, one can look at Fig.\ref{fig5}(a) {(often called a $\mathcal{PT}$ phase diagram\cite{pt})} which shows that topological end states with imaginary eigenvalues begin to emerge soon after $\gamma$ is increased from a zero value. We also notice that the eigenvalues near $\Delta/t\rightarrow 0$ remain purely real. However, $\mathcal{SPT~BP}$ appears for larger $\gamma~(\sim3\times10^{-2})$ values with the presence of end states with imaginary energy even at $\Delta=0$ [Fig.\ref{fig4}(b)]. Contrarily, the bulk states remain real until $\gamma\sim5.1\times10^{-2}$ when complex-valued gapped bulk modes start appearing at the TQPT point of $|\Delta/t|=\sqrt{2}$ [Fig.\ref{fig4}(c)]. With further increase in $\gamma$ say at $\gamma=7\times10^{-2}$ [Fig.\ref{fig4}(d)],  new complex bulk modes start sprouting also in the vicinity of $|\Delta/t|\rightarrow 0$.

Unlike the $\theta=\pi$ case, this case admits 2 new in-gap states\cite{mandal} which show real eigenvalues {in the Hermitian limit ($\gamma=0$).} Interestingly, the in-gap modes start acquiring imaginary eigenvalues {for much higher values of $\gamma$ compared to both end states and other bulk states (see Fig.\ref{fig4}-\ref{fig5}). Notice that they possess $Im[E]\neq0$ for $\gamma\sim1.6$ at $|\Delta|/t\le2.5\times10^{-1}$ (Fig.\ref{fig4}(h)) but lack any imaginary eigenvalue for smaller $\gamma$ values.}

For $\gamma=2$ [see Fig.\ref{fig4}(i)] the bulk modes gain purely imaginary eigenvalues near $\Delta/t\rightarrow 0$ and in-gap states become imaginary conjugated pairs for $|\Delta/t|\le7.32\times10^{-1}$ [Fig.\ref{fig4}(i)]. We find that from this critical value $\gamma_c=2$ onwards, an increasing $\gamma$ results in further stretched ranges of $|\Delta|/t$ for which purely imaginary bulk modes as well as in-gap modes are obtained. Fixing $\Delta/t$ value at $7.32\times10^{-1}$, how the imaginary spectrum of the bulk states, in-gap pairs or end-state pairs changes with $\gamma$ can be noticed from Fig.\ref{fig5}(a). For $\gamma>2$, {both the in-gap and other bulk modes} have purely imaginary eigenvalues which lasts for larger ranges of $\Delta/t$ that eventually (with an increase in $\gamma$) covers whole TNP [for $\gamma\sim 2.8$] and then go beyond [Fig.\ref{fig4}(k)].

{Thus to summarize the bulk spectrum,} the bulk modes for weak $\gamma$ [Fig.\ref{fig4}(a)-(d)] don't support any imaginary eigenvalue except at the gap-closing points. For intermediate values of $\gamma$, one gets complex bulk mode eigenvalues [Fig.\ref{fig4}(e)-(i)] While for large $\gamma$, the bulk states hold purely imaginary eigenvalues for small $|\Delta|$[Fig.\ref{fig4}(j)-(k)].

\subsubsection{{Trivial Phase}}

Now we turn to the {trivial phase} region i.e., the regime of $|\Delta/t|>\sqrt{2}$. Again, real bulk bands are found for smaller $\gamma$ values. With larger $\gamma$, for example for $\gamma\gtrsim5.0\times10^{-2}$ [Fig.\ref{fig4}(c)], both TNP and TTP close to the TQPT point feature the bulk modes with conjugated imaginary eigenvalues. Within TTP, the energy spectra is entirely real below such critical $\gamma$ turning this into a $\mathcal{SPT~BT}$ point.

Within TTP, the energy of the in-gap states, {that were real in the Hermitian limit,} start becoming imaginary starting from TQPT point beyond $\gamma\sim2.7$ [see Fig.\ref{fig4}(k)] {for a range of $\Delta$ that covers more and more TTP space with further increase in $\gamma$.} One can notice such behavior in the energy spectrum plot of Fig.\ref{fig5}(d) corresponding to $\Delta/t=1.456$. For $\gamma\gtrsim2$, one can also observe that a larger $\gamma$ pushes the critical $|\Delta|$ to higher values above which nonzero real components of the bulk modes are obtained within TTP.
\begin{figure}[htp]
   \vskip -.1 in
   \begin{picture}(100,100)
     \put(-100,0){
       \includegraphics[width=.55\linewidth]{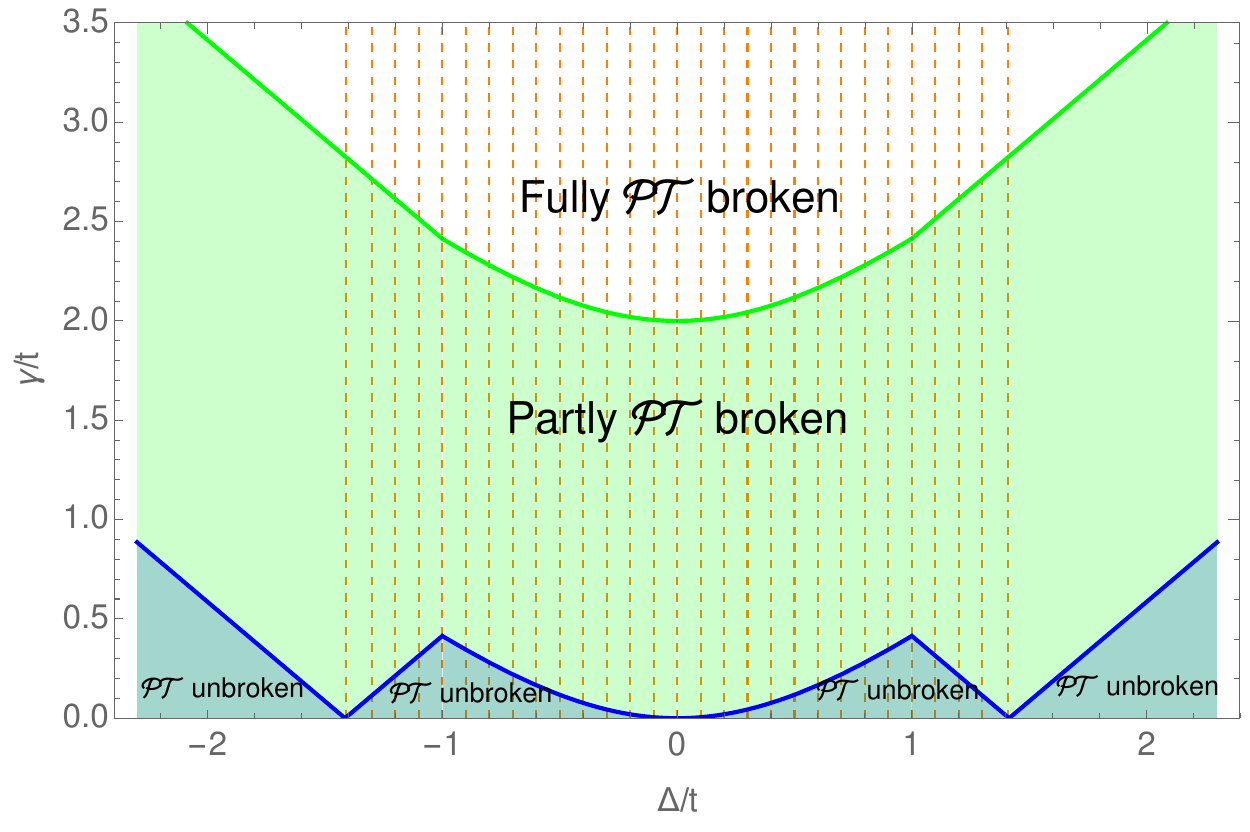}\includegraphics[width=.55\linewidth]{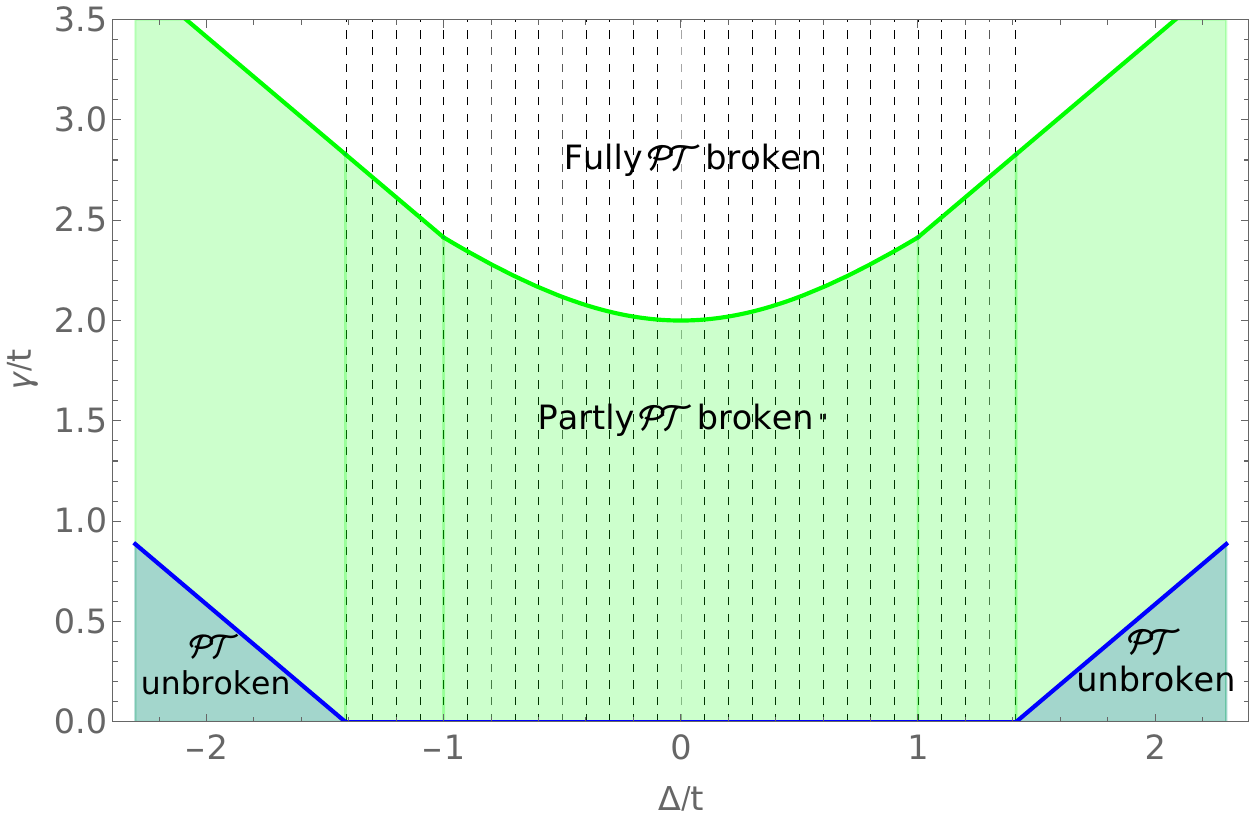}}
     \put(-80,72){(a)}
     \put(55,72){(b)}
   \end{picture} 
  %\vskip -0.1 in
\caption{Phase diagram of a {(a)} periodic and {(b) open chain} in a $\gamma-\Delta$ plane for $\theta=\pi/2$. Generally for small $\gamma$ one gets $\mathcal{PT}$ symmetric states which then first partly (for a few momentum states of the BZ) and then finally fully broken as $\gamma$ is gradually increased. Also, TNP can be identified from the vertically striped regime.}
\label{ph-diag2}
\end{figure}

Interestingly, a NH SSH model with imaginary potentials only at the boundaries can result in bifurcation of the imaginary eigenvalues for large $\gamma$ values\cite{zhu}.
Nevertheless, very close to phase transition in a part of the topological phase shows a $\mathcal{SPT~BT}$ ($\mathcal{SPT~BP}$) at (followed by) the $\mathcal{PT}$ symmetry breaking point as long $\gamma~=(>)~2.7\times10^{-3}$ rather than considering very weak value ($\sim 0$) (as discussed in the previous case) [Fig.\ref{fig5}(b)]. Thus $\gamma_{ep}$ becomes $2.7\times10^{-3}$. Though $\Delta/t=0$ is not TQPT point for the Hermitian system\cite{mandal}, one can visualize the occurrence of $\mathcal{PT}$ phase transition near $\Delta/t=0$ in the same fashion as found in the vicinity of TQPT. Fig.\ref{fig5}(c) depicts that imaginary eigenvalues begin to emerge at the TQPT point at $\gamma=~1.7\times 10^{-1}$.

\begin{figure}[htp]
   \vskip 0 in
   \includegraphics[width=\linewidth]{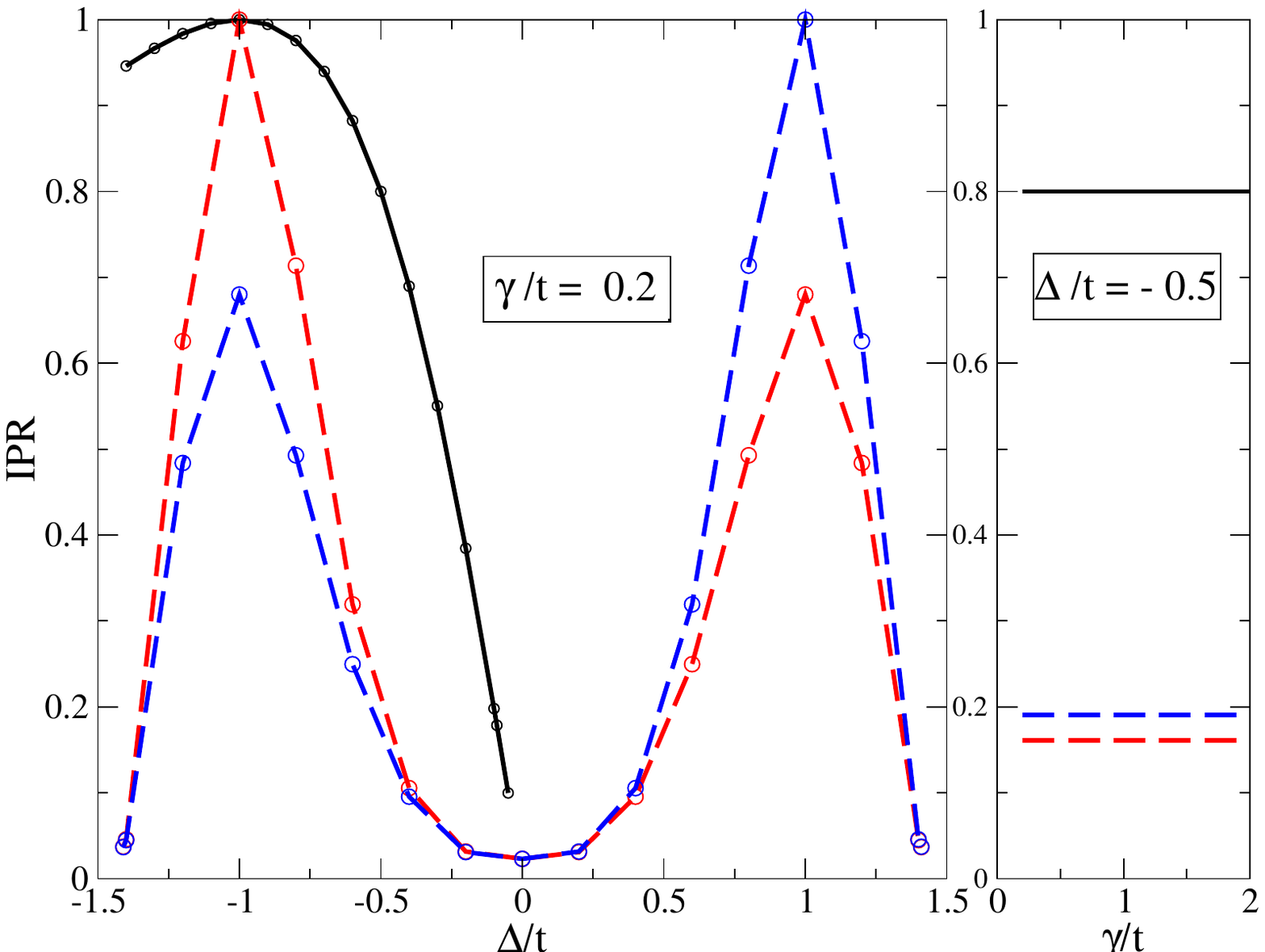}
   \caption{IPR of the NH ZES or end states for $\theta=\pi$ (solid) and $\pi/2$ (dashed) for different values of $\Delta$ and $\gamma$.}
   \label{ipr2}
\end{figure}
Overall, we show the phase diagram of the NH SSH chain {for $\theta=\pi/2$ in Fig.\ref{ph-diag2} under ((a)) PBC and ((b)) OBC, highlighting all of TNP, TTP as well as $\mathcal{PT}$ symmetric and the $\mathcal{PT}$ partially and fully broken regions. Like in the $\theta=\pi$ case, here also the TQPT point\cite{cmnt} comes out to be independent of $\gamma$ values\cite{sudin}. Under PBC, all the bulk modes remain real for small $\gamma$ values, apart from any gap-closing point ($i.e.,$ TQPT and $\Delta=0$ point). Contrarily under OBC, the $\mathcal{PT}$ symmetric region practically vanishes within TNP as the end states soon become imaginary as $\gamma$ is switched on.}
We should note here that in Ref.\cite{sudin} a modified definition of winding number is considered in order to respect the $\mathcal{PT}$ symmetries and it leads to a winding number giving unity (fraction or zero) for $\mathcal{PT}$ unbroken (partly broken or fully broken) phases.

\begin{figure}[htp]
   \includegraphics[width=1.15\linewidth]{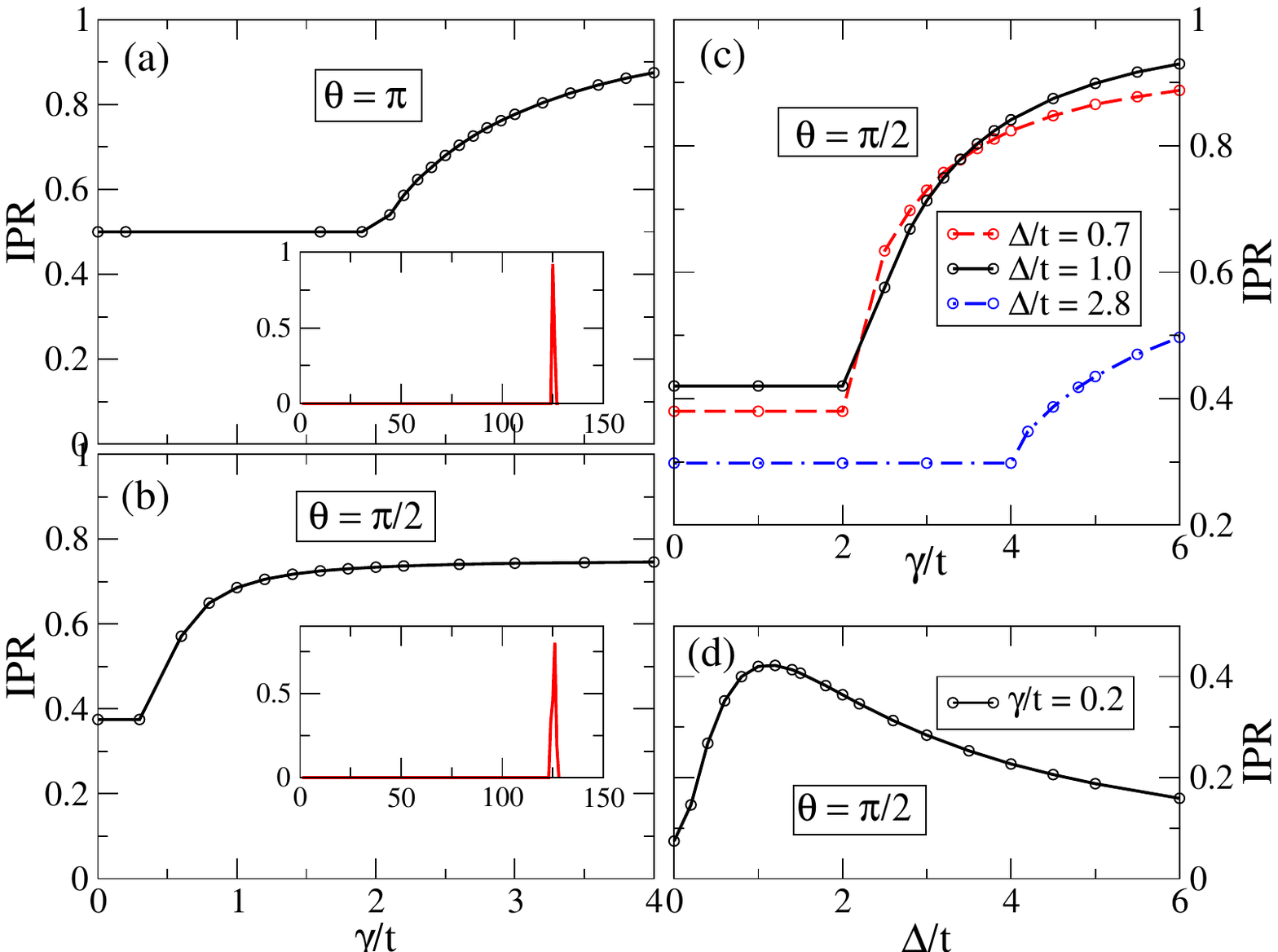}
%   \vskip 1.5 in
   \caption{{IPR of the degenerate bulk states at $\Delta/t=1$ for (a) $\theta=\pi$ and (b) $\pi/2$ for different values of $\gamma$ (with L=128). Insets shows such states localized at one end. (c) and (d) show the IPR plots of in-gap states with respect to $\gamma$ and $\Delta$ respectively.}}
   \label{ipr3}
   \end{figure}

    We calculate the inverse participation ratio (IPR)\cite{roy} of the end states  as estimators of their the localization within the topological phase for different $\gamma$ values. {Though in general we don't see any $\gamma$ dependence,} these IPRs do have a $\Delta$ dependence that loses strength fast (indicating advent of extended states) as one alters $\Delta$ towards isotropic hopping limit of $\Delta=0$ within TQPT (see Fig.\ref{ipr2}). Notice that for $\theta=\pi/2$, the end states of the two ends have different IPRs due to the difference of hopping distribution seen from the two ends. Both the phase diagrams and the IPR plots indicate a specialty of the point $|\Delta/t|=1$ {which can be tagged as maximally dimerized limit where the factor of $(t^2-\Delta^2)$ appearing in both Eq.\ref{6} and Eq.\ref{8} vanishes. Lacking continuous hopping pathways, these conditions, irrespective of whether within TTP or TNP, always produce degenerate eigenstates localized around separate nearest-neighbor bonds. In Fig.\ref{ipr3}(a),(b) we show the variation of the bulk state IPRs with $\gamma$ at $\Delta/t=1$ for both $\theta=\pi$ and $\pi/2$. Out of them, the ones localized at edges can be counted in for the display of skin effects. Notice that the increase in localization with $\gamma$ occurs only in partly or fully $\mathcal{PT}$-broken phases. In-gap states also show an increase in localization with $\gamma$ (Fig.\ref{ipr3}(c)) and contribute to the skin effect while they show an extended nature at $\Delta\rightarrow0$ and away from the maximally dimerized limit of $|\Delta/t|=1$ (Fig.\ref{ipr3}(d)).}

\subsection{Case of $\theta=\pi/4$}

One obtains a $8\times8$ Bloch Hamiltonian $H_{\mathcal{PT}}(k)$ as
\begin{widetext}
\begin{equation}\label{9}
H_{\mathcal{PT}}(k) =
\begin{pmatrix}
i\gamma & (t+\Delta) & 0 & 0 & 0 & 0 & 0 & (t+\frac{\Delta}{\sqrt{2}})e^{-8ik} \\
(t+\Delta) & -i\gamma & (t+\frac{\Delta}{\sqrt{2}}) & 0 & 0 & 0 & 0 & 0 \\
0 & (t+\frac{\Delta}{\sqrt{2}}) & i\gamma & t & 0 & 0 & 0 & 0 \\
0 & 0 & t & -i\gamma & (t-\frac{\Delta}{\sqrt{2}}) & 0 & 0 & 0 \\
0 & 0 & 0 & (t-\frac{\Delta}{\sqrt{2}}) & i\gamma & (t-\Delta) & 0 & 0\\
0 & 0 & 0 & 0 & (t-\Delta) & -i\gamma & (t-\frac{\Delta}{\sqrt{2}} & 0 \\
0 & 0 & 0 & 0 & 0 & (t-\frac{\Delta}{\sqrt{2}}) & i\gamma & t \\
(t+\frac{\Delta}{\sqrt{2}})e^{8ik} & 0 & 0 & 0 & 0 & 0 & t & -i\gamma  \\
\end{pmatrix}
\end{equation}
\end{widetext}

\begin{figure}
    \vskip -.6 in
   \begin{picture}(100,100)
     \put(-70,0){
  \includegraphics[width=.45\linewidth,height= .8 in]{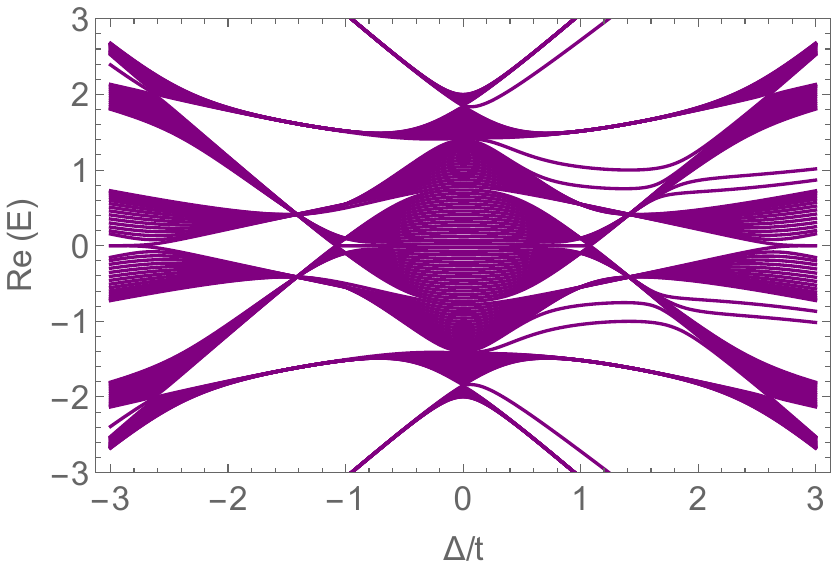}\hskip .2 in
  \includegraphics[width=.45\linewidth,height= .8 in]{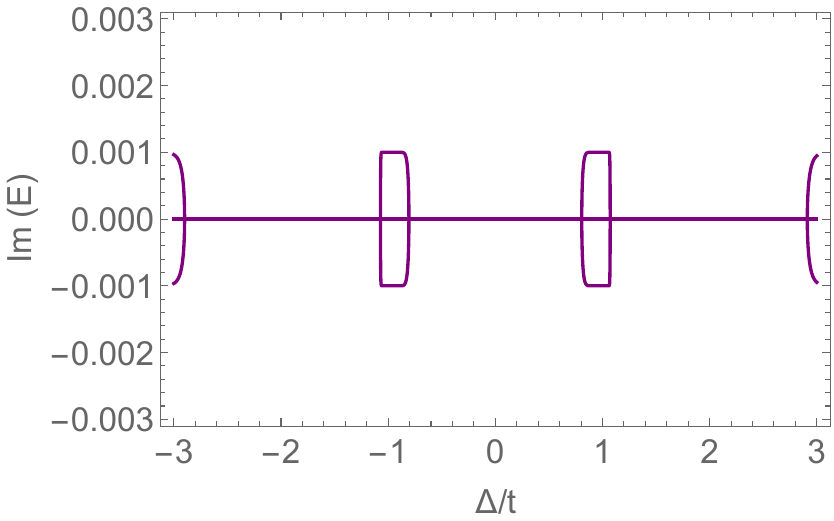}}
     \put(47,30){(a)}
   \end{picture}\\
   \vskip -.6 in
   \begin{picture}(100,100)
     \put(-70,0){
   \includegraphics[width=.45\linewidth,height= .8 in]{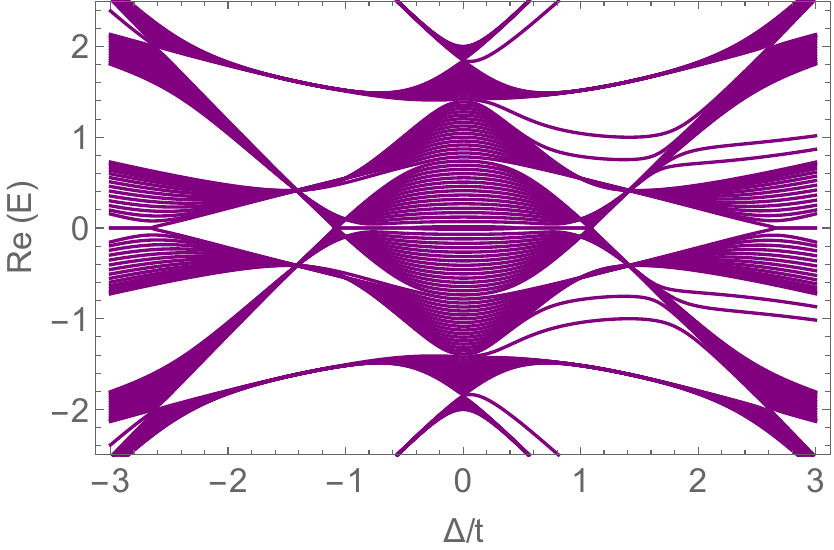}\hskip .2 in
   \includegraphics[width=.45\linewidth,height= .8 in]{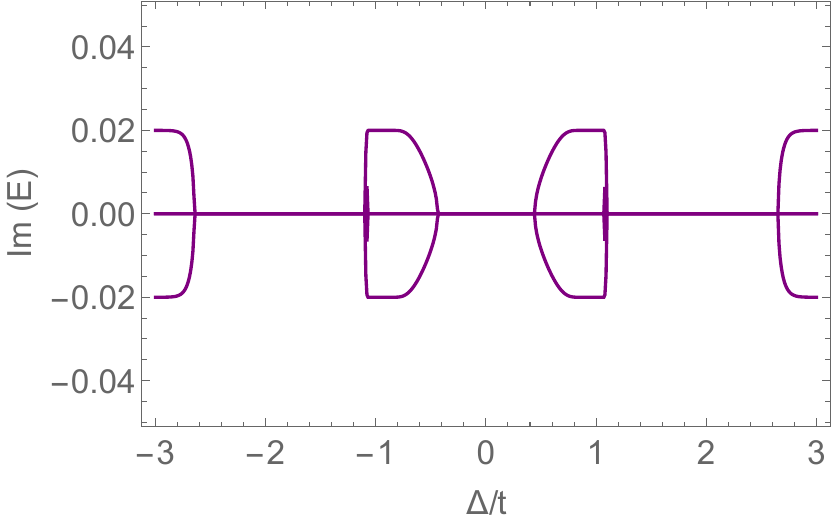}}
   \put(47,30){(b)}
   \end{picture}\\
    \vskip -.6 in
  \begin{picture}(100,100)
     \put(-70,0){
   \includegraphics[width=.45\linewidth,height= .8 in]{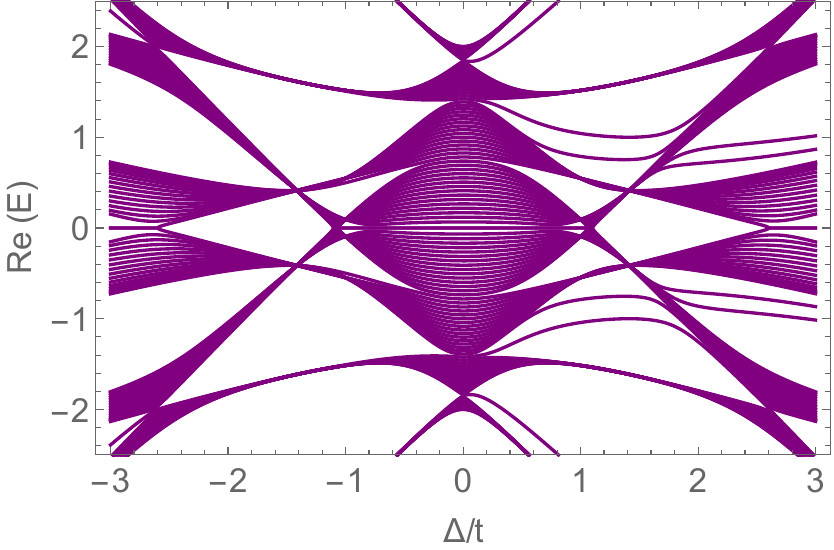}\hskip .2 in
   \includegraphics[width=.45\linewidth,height= .8 in]{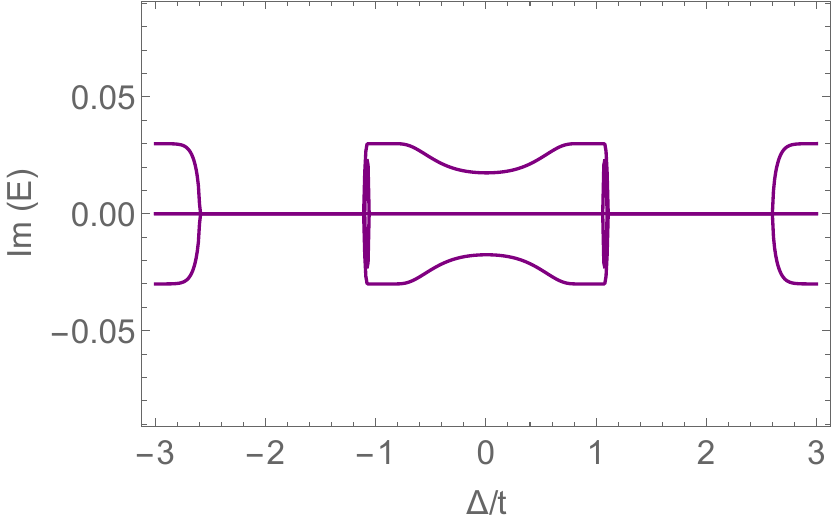}}
   \put(47,30){(c)}
   \end{picture}\\
    \vskip -.6 in \begin{picture}(100,100)
     \put(-70,0){
   \includegraphics[width=.45\linewidth,height= .8 in]{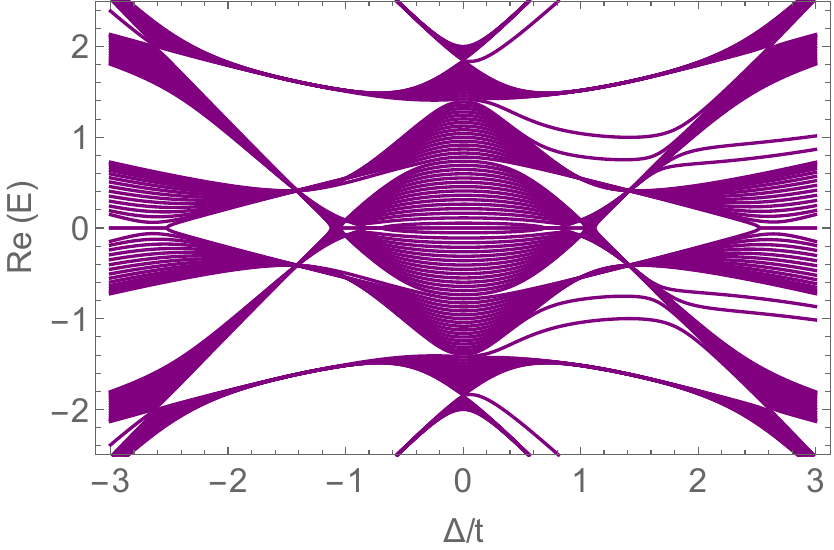}\hskip .2 in
   \includegraphics[width=.45\linewidth,height= .8 in]{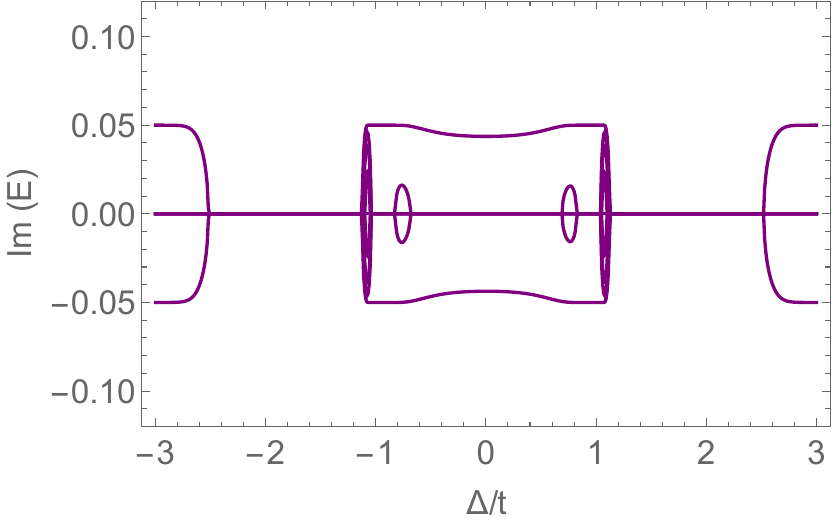}}
   \put(47,30){(d)}
   \end{picture}\\
    \vskip -.6 in
 \begin{picture}(100,100)
     \put(-70,0){
   \includegraphics[width=.45\linewidth,height= .8 in]{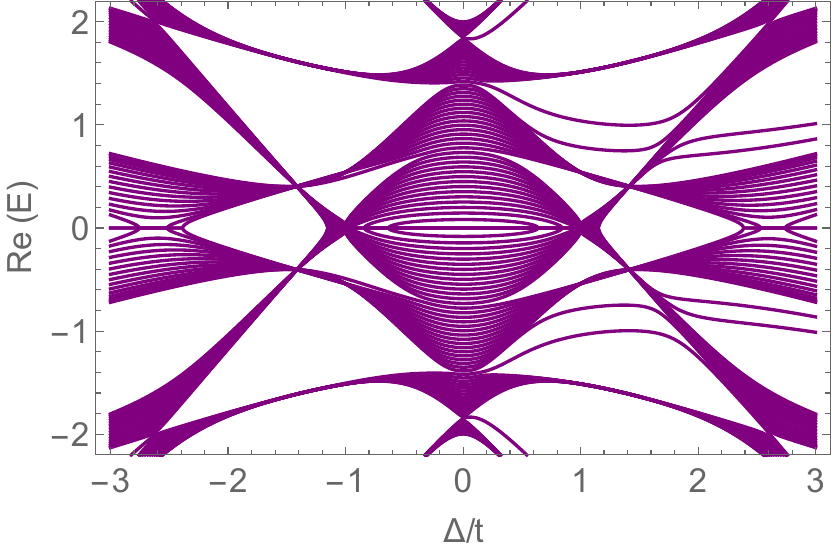}\hskip .2 in
   \includegraphics[width=.45\linewidth,height= .8 in]{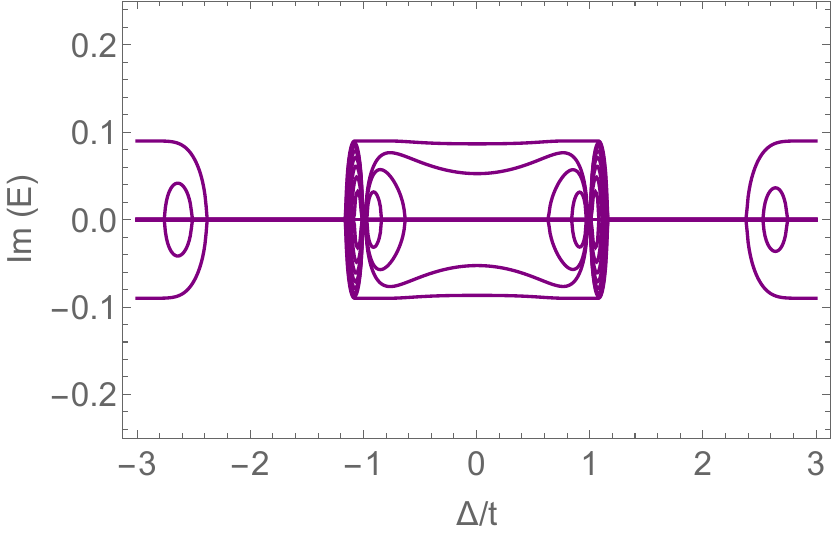}}
   \put(47,30){(e)}
   \end{picture}\\
    \vskip -.6 in  \begin{picture}(100,100)
     \put(-70,0){
   \includegraphics[width=.45\linewidth,height= .8 in]{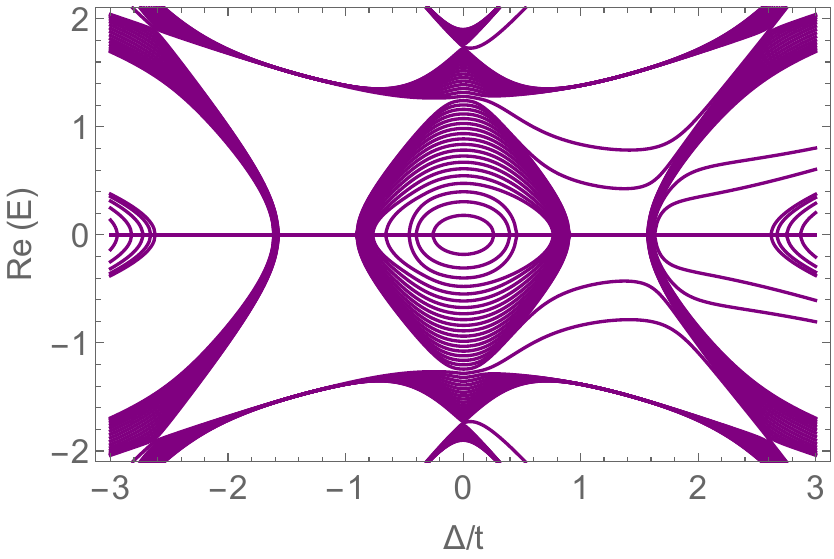}\hskip .2 in
   \includegraphics[width=.45\linewidth,height= .8 in]{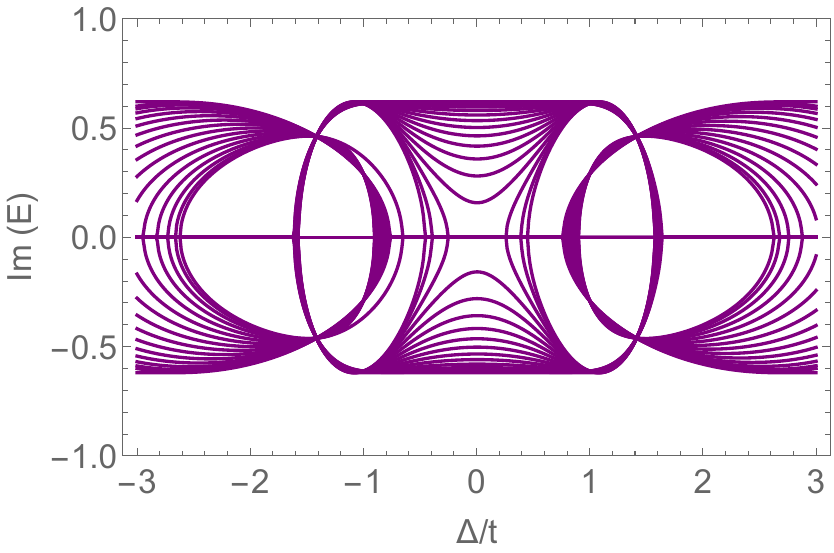}}
   \put(47,30){(f)}
    \end{picture}
   \vskip -.6 in
   \begin{picture}(100,100)
     \put(-70,0){
  \includegraphics[width=.45\linewidth,height= .8 in]{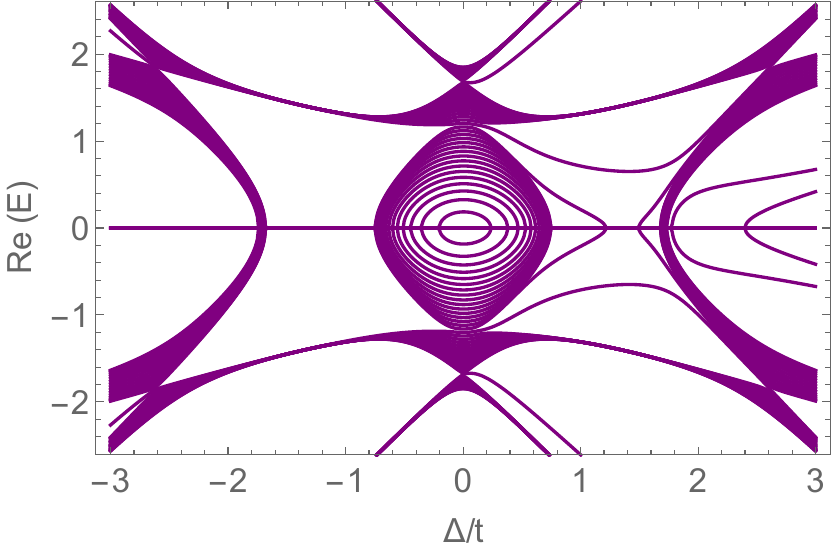}\hskip .2 in
  \includegraphics[width=.45\linewidth,height= .8 in]{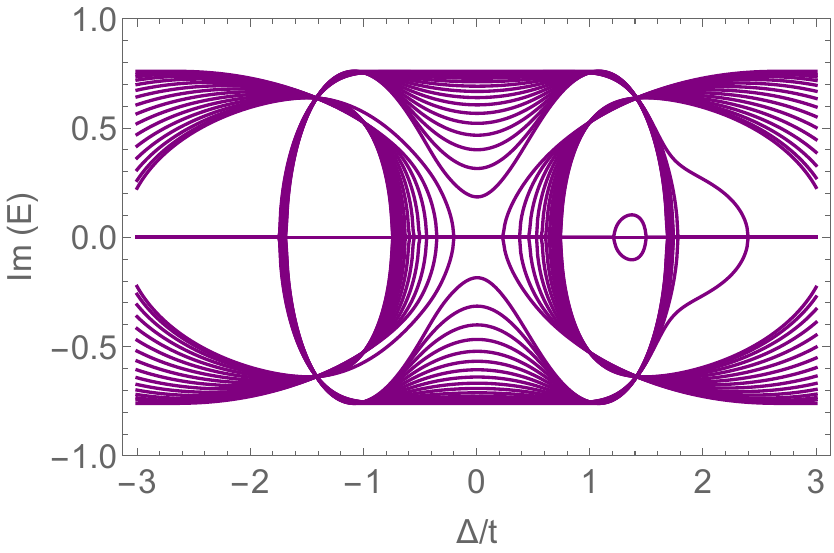}}
     \put(47,30){(g)}
   \end{picture}\\
   \vskip -.6 in
  \begin{picture}(100,100)
     \put(-70,0){
   \includegraphics[width=.45\linewidth,height= .8 in]{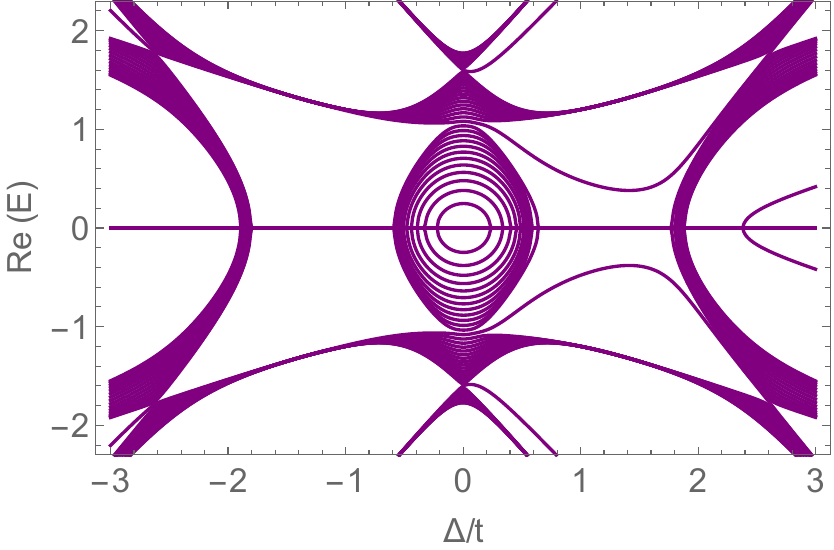}\hskip .2 in
   \includegraphics[width=.45\linewidth,height= .8 in]{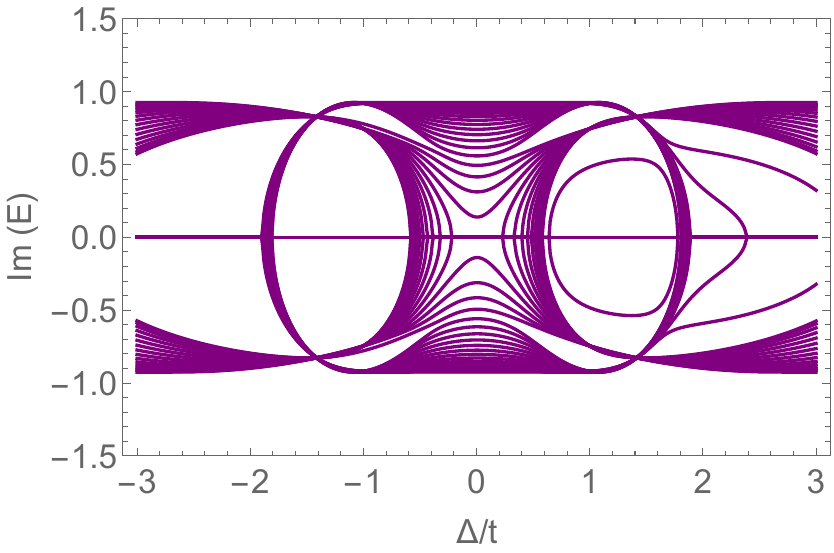}}
   \put(47,30){(h)}
   \end{picture}\\
    \vskip -.6 in \begin{picture}(100,100)
     \put(-70,0){
   \includegraphics[width=.45\linewidth,height= .8 in]{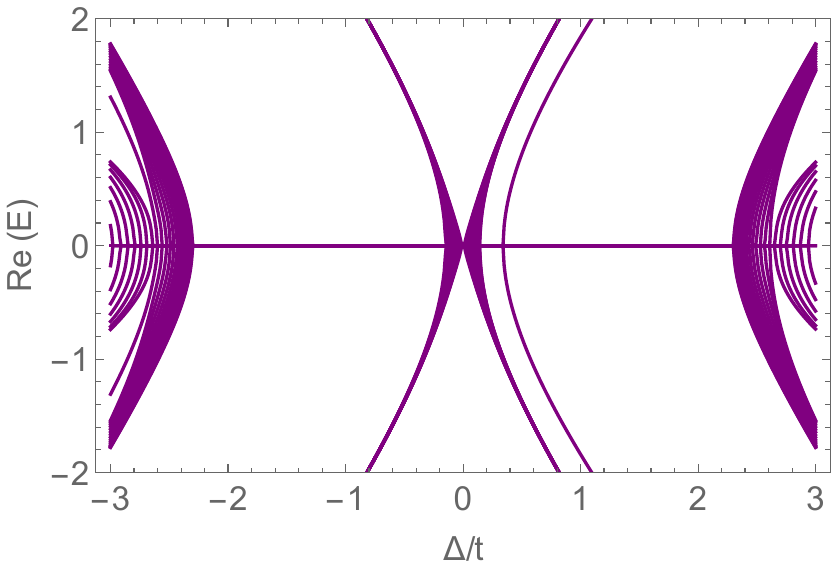}\hskip .2 in
   \includegraphics[width=.45\linewidth,height= .8 in]{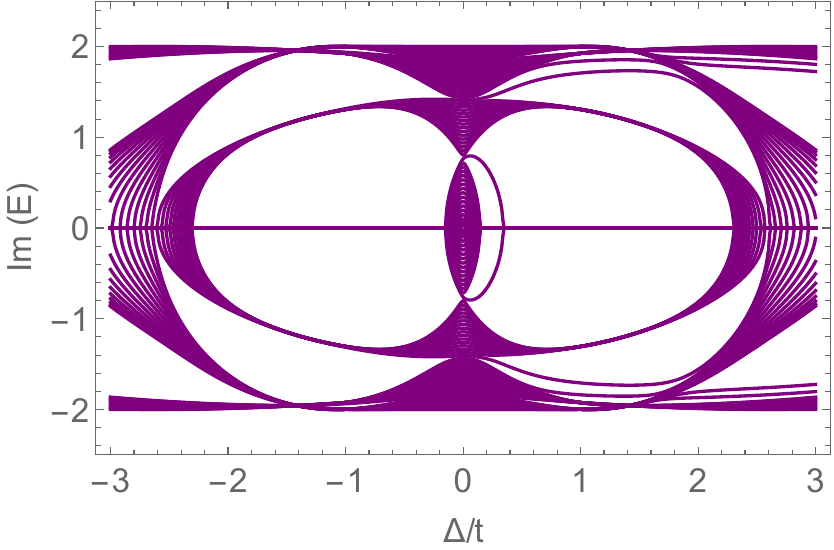}}
   \put(47,30){(i)}
   \end{picture}\\
    \vskip -.6 in
 \begin{picture}(100,100)
     \put(-70,0){
   \includegraphics[width=.45\linewidth,height= .8 in]{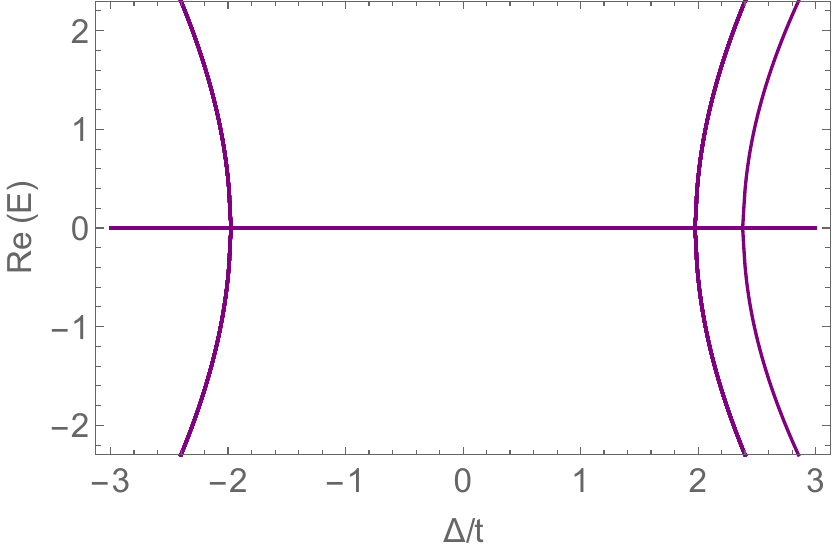}\hskip .2 in
   \includegraphics[width=.45\linewidth,height= .8 in]{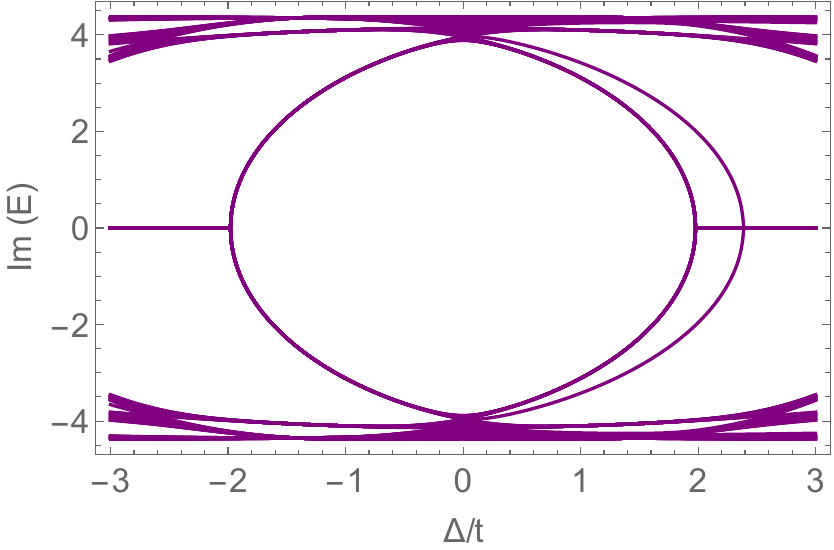}}
   \put(47,30){(j)}
    \end{picture}
  %\vskip -0.1 in
\caption{Numerical spectra of the {\bf real} (left) and {\bf imaginary} (right) part of the energy eigenvalues of an {\bf open chain} for the $\mathcal{PT}$-symmetric NH SSH model with $\Delta/t$ ($\in[-3,~3]$) with $t=1$,~$L=128$ and for various $\gamma$~=~ (a)$1\times10^{-3}$;~(b)$2\times 10^{-2}$;~(c)$3\times10^{-2}$;~(d)$5\times10^{-2}$;~(e)$9\times10^{-2}$;~(f)$6.2\times10^{-1}$;~(g)$7.6\times10^{-1}$;~(h)$9.25\times10^{-1}$;~(i)2;~(j)4.37.} 
\label{fig6}
\end{figure}

\begin{figure}
   \vskip -.4 in
   \begin{picture}(100,100)
     \put(-70,0){
  \includegraphics[width=.48\linewidth]{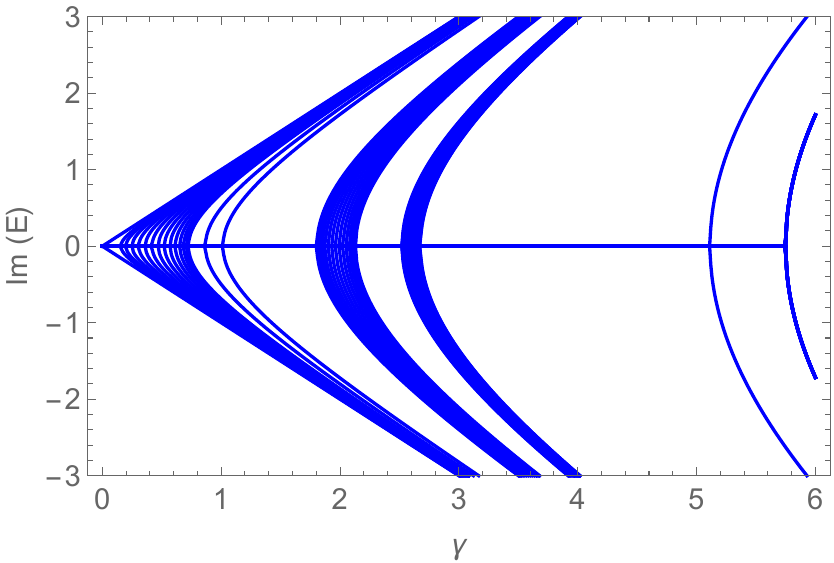}
  \includegraphics[width=.48\linewidth]{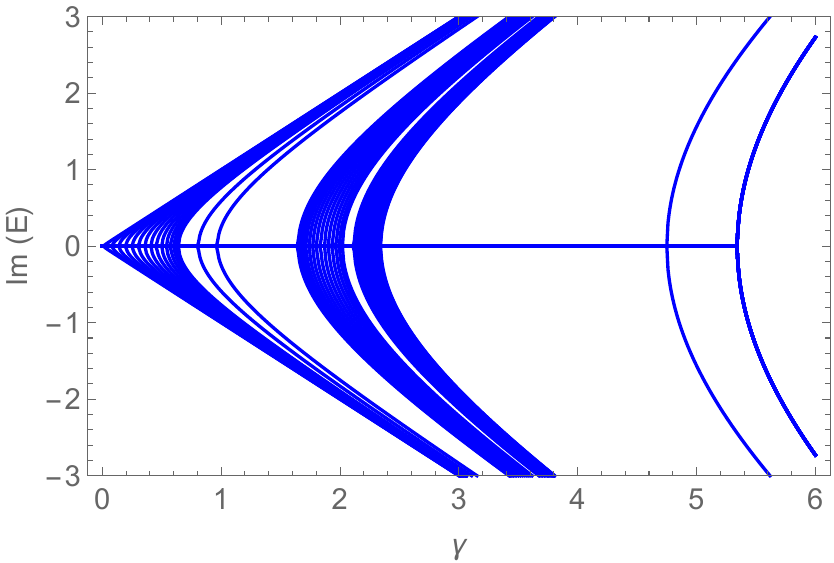}}
     \put(-50,68){(a)}
     \put(75,68){(b)}
   \end{picture}\\
   \vskip -.2 in
   \begin{picture}(100,100)
     \put(-70,0){
   \includegraphics[width=.48\linewidth]{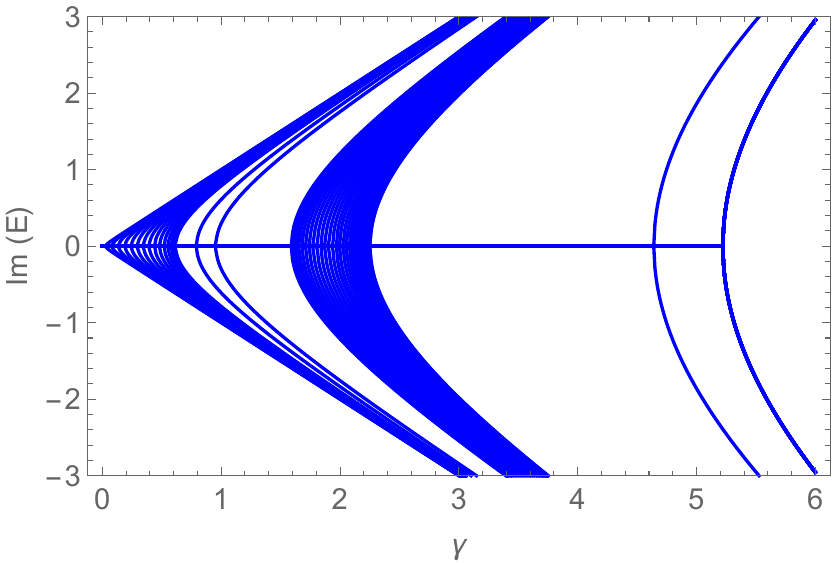}
   \includegraphics[width=.48\linewidth]{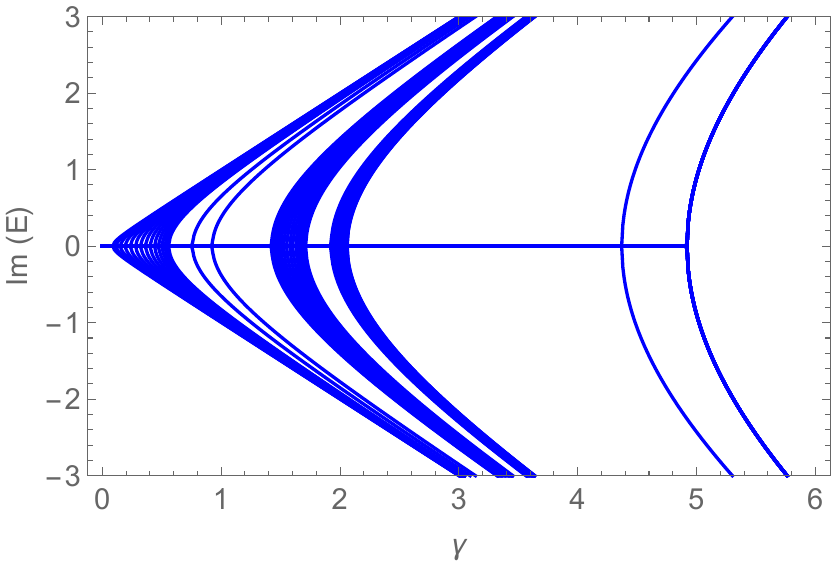}}
   \put(-50,68){(c)}
   \put(75,68){(d)}
    \end{picture} 
  %\vskip -0.1 in
\caption{Numerical spectra of the {\bf imaginary part} of the eigenvalue under {\bf OBC} with respect to $\gamma$ with $t=1$,~$L=128$ and $\Delta/t$~=~(a)~3 (system in the TNP), (b)~2.7 (close to phase boundary in a part of the TNP),(c)~2.61 (at the phase boundary) and (d)2.387 (system in the TTP). The inset shows the appropriate position of $\mathcal{SPT~BT}$ point.}
\label{fig7}
\end{figure}

The real and imaginary part of the energy eigenvalues with $\Delta/t$ for this case is presented in Fig.\ref{fig6}. Spectral features are similar to the previous two $\theta$ values considered in many aspects. Topological end states can also be visualized from the $\mathcal{PT}$ phase diagram [Fig.\ref{fig7}]. The region of $\Delta/t\rightarrow 0$ no longer remains $\mathcal{PT}$-symmetric when $\gamma>2\times10^{-2}$. One can notice the imaginary bulk modes to appear near the TQPT point of $|\Delta/t|=\sqrt{2(2-\sqrt{2})}$ for $\gamma=2\times10^{-2}$ or of $|\Delta/t|=\sqrt{2(2+\sqrt{2})}$ for $\gamma=9\times10^{-2}$ [see Fig.\ref{fig6}(c),(e)]. This case includes six new in-gap states that illustrate entirely real eigenvalues when $0<\gamma\lesssim9\times10^{-2}$. For critical $\gamma=\gamma_c=2$, the eigenvalues of bulk states become purely imaginary in the vicinity of $\Delta/t\rightarrow 0$ while the in-gap states become imaginary at $\Delta/t\lesssim 3.5\times10^{-1}$. However, a larger $\gamma>2$ drives a large region about $\Delta/t\rightarrow 0$ to become purely imaginary. In Fig.\ref{fig6}(f), two (among six) in-gap modes become complex within topological phase, specifically near the phase boundary $\Delta/t=\sqrt{2(2+\sqrt{2})}$,  for $\gamma=7.97\times10^{-1}$.

Within {trivial phase}, one can notice only the existence of real bulk modes when $\gamma$ is weak i.e., $\gamma=1\times10^{-3}$. For $\gamma\gtrsim2\times 10^{-2}$ [Fig.\ref{fig6}(b)], complex eigenvalues of the bulk start to emerge gradually growing around the phase boundary of $|\Delta/t|=\sqrt{2(2-\sqrt{2})}\simeq1.082$. On the other hand, complex eigenvalues appear near the other TQPT point of $|\Delta/t|=\sqrt{2(2+\sqrt{2})}\simeq2.613$ for $\gamma\gtrsim9\times10^{-2}$ [Fig.\ref{fig6}(e)]. Thus one obtains two critical values of $\gamma_{c_{1}}\sim2\times10^{-2}$ and $\gamma_{c_{2}}\sim9\times10^{-2}$. For instance, the imaginary spectrum against $\gamma$ with $\Delta/t=2.387$ is plotted in Fig.\ref{fig7}(d) to see the changes in the spectrum with $\gamma$. It can be seen from the plot that the spectrum has purely real eigenvalues when $\gamma_{c_{2}}$ and beyond this the system enters into $\mathcal{SPT~BP}$. More interestingly, here three complex pairs of in-gap modes start appearing at the same $|\Delta/t|$ (= 2.387) but for three typical values of $\gamma$ as $7.6\times10^{-1},~9.25\times10^{-1},~4.37$. [see Fig.\ref{fig6}(g),(i),(k)]. How these different values of $\gamma$ can result in the emergence of complex pairs in the spectrum for in-gap states is depicted in the $\mathcal{PT}$ phase diagram as shown in Fig.\ref{fig7}(d). Additionally, Fig.\ref{fig7}(a),(b) displays the similar trend of occurring the $\mathcal{SPT~BT}$ and $\mathcal{SPT~BP}$ in the TNP. We find the complex eigenvalues would emerge at $\gamma=2.2\times10^{-2}$ at the TQPT point of $\Delta/t=\sqrt{2(2+\sqrt{2})}$ [Fig.\ref{fig7}(c)].

The above analysis for all commensurate $\theta$ values indicates that the periodic modulated SSH system undergoes a $\mathcal{SPT~BT}$ in the topological phase at a critical $\gamma_{ep}$ (known as EP) close to a topological transition point. But $\gamma_{ep}\rightarrow0$ as one looks deep within TNP (see Fig.\ref{fig3}(a)), resulting in $\mathcal{SPT~BP}$ for any arbitrary nonzero $\gamma$ values. Contrarily, there exist only real eigenvalues in the TTP for some $\gamma<\gamma_c$, above which all bulk state eigenvalues become complex leading to the occurrence of $\mathcal{SPT~BP}$.

\subsection{{Features} of Complex End, In-Gap and Bulk States for $\theta=~\pi/2,~\pi/4$}

\begin{figure}
   \vskip -.4 in
   \begin{picture}(100,100)
     \put(-85,0){
  \includegraphics[width=.34\linewidth,height= 1 in]{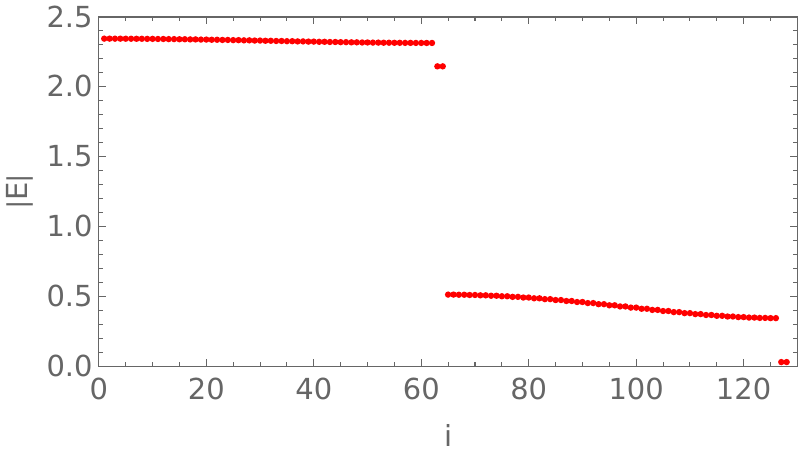}
  \includegraphics[width=.34\linewidth,height= 1 in]{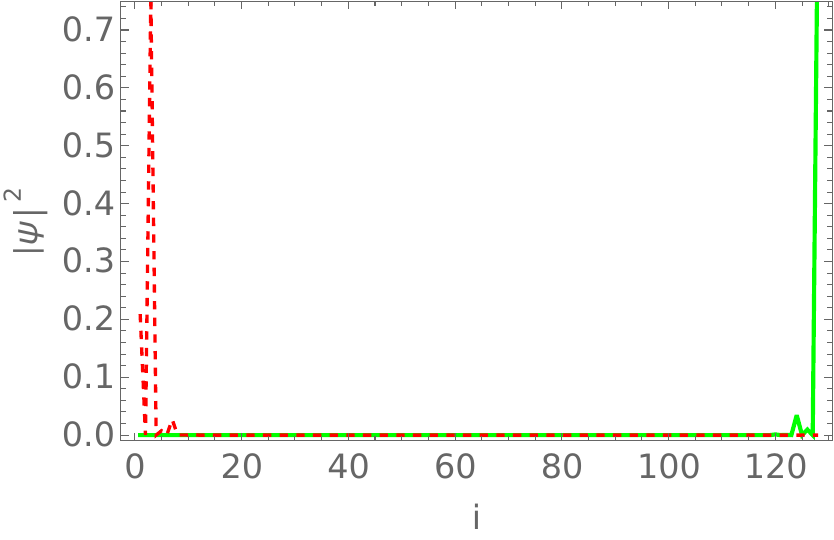}
   \includegraphics[width=.34\linewidth,height= 1 in]{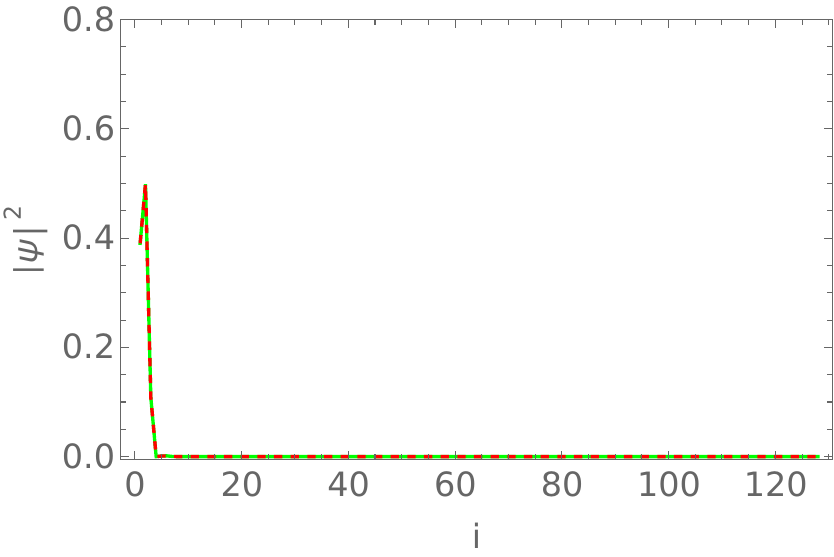}}
     \put(-15,55){(a)}
     \put(70,55){(b)}
     \put(160,55){(c)}
   \end{picture}\\
   \vskip -.3 in
   \begin{picture}(100,100)
     \put(-85,0){
   \includegraphics[width=.34\linewidth,height= 1 in]{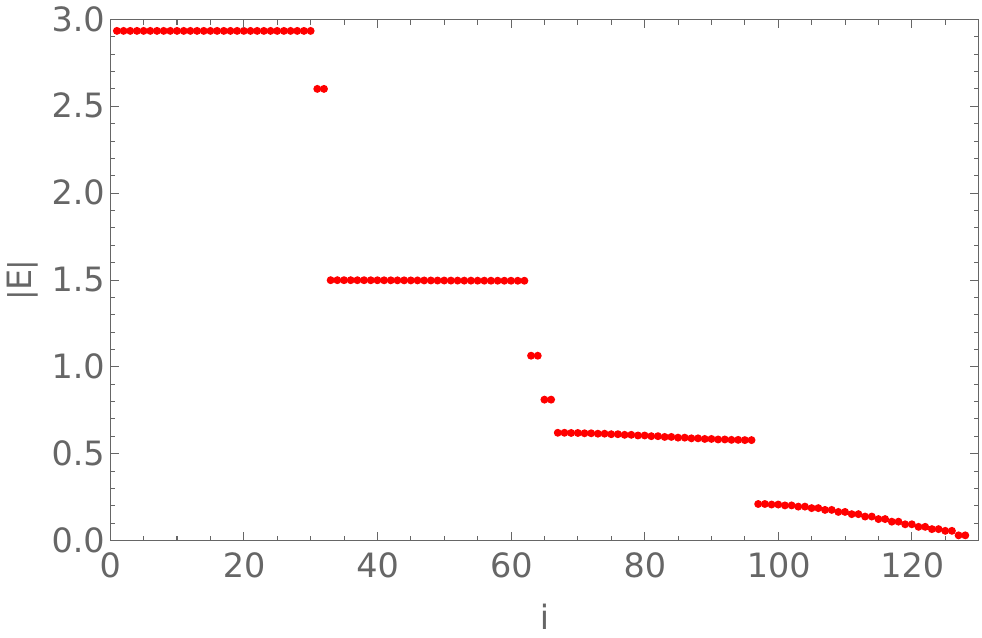}
   \includegraphics[width=.34\linewidth,height= 1 in]{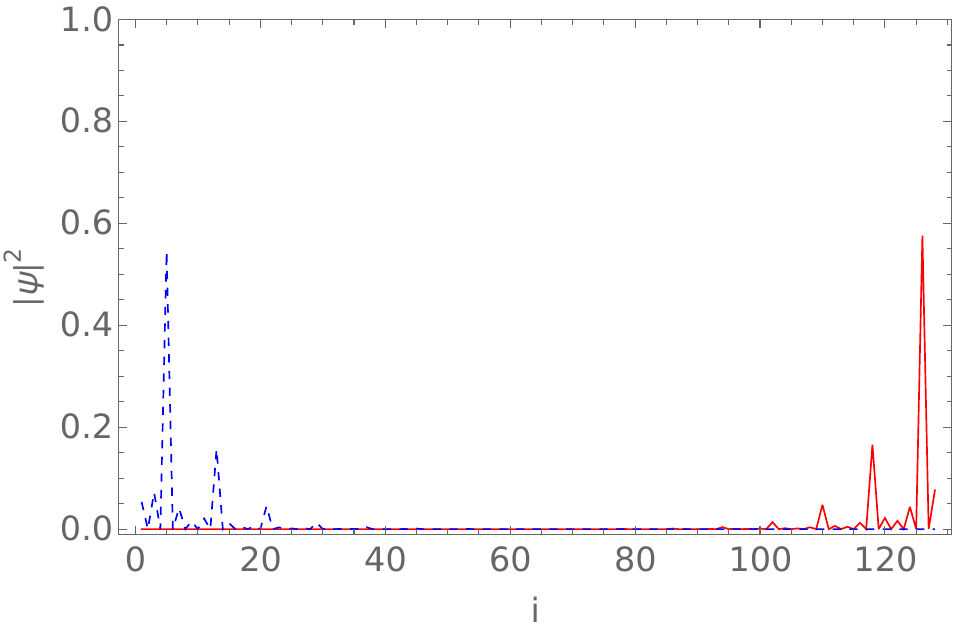}
    \includegraphics[width=.34\linewidth,height= 1 in]{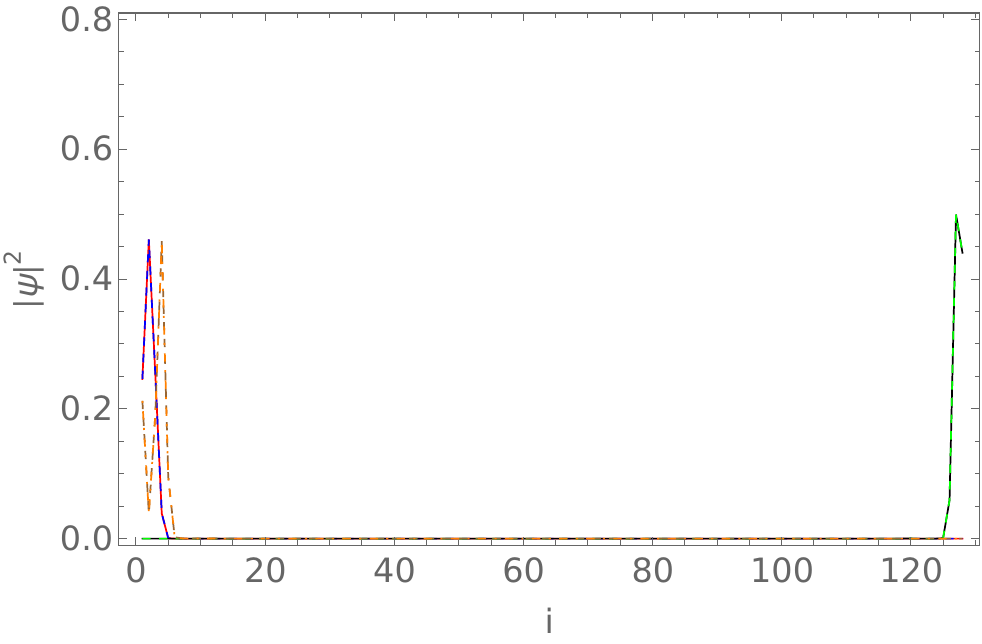}}
   \put(-15,55){(d)}
   \put(70,55){(e)}
    \put(160,55){(f)}
    \end{picture} 
  %\vskip -0.1 in
\caption{{Magnitude of energy eigenvalues for all the eigenstates corresponding to  (a) $\theta=\pi/2$ and (d) $\pi/4$. Probability density of end states (b,e) and in-gap states (c,f) along the chain of length L=128 at $\Delta/t=0.9$, $\gamma/t=0.03$ and for $\theta=\pi/2$ and $\pi/4$ respectively.}}
\label{fig8}
\end{figure}

{Now we make a comparative analysis of the complex spectrum of the hopping modulated NH SSH chain, namely for $\theta=\pi/2$ and $\pi/4$. The absolute energy eigenvalues clearly show the separately positioned end states (within TNP) and in-gap states between the continuum of bulk states in the spectrum (see Fig.\ref{fig8}(a),(d)).}

{For $\theta=\pi/2$ there is a single tiny TNP region about $\Delta=0$ whereas for $\theta=\pi/4$ we find two additional semi-infinite TNP regions for $|\Delta/t|>\sqrt{2(2+\sqrt{2})}$.
As we keep increasing $\gamma$ within TNP, we have seen that for $\theta=\pi/2$, first NH ZES become imaginary at $\gamma_{ep}$. Then for a higher value of $\gamma$ bulk modes start becoming complex and at an even larger $\gamma$, the in-gap states (2 in number) turn from real to imaginary (of-course at $\gamma$ values dependent on $\Delta$ chosen). Similar trend is found for $\theta=\pi/4$ but there are differences as well. Unlike in the $\theta=\pi/2$ case, the spectra is not symmetric about $\Delta=0$. There are 3 pairs of in-gap states which show different critical $\gamma$ for their conversion of eigenvalues from real to imaginary. Moreover, the critical $\Delta$ where such conversion occurs first is at a TQPT (unlike at $\Delta=0$ in the $\theta=\pi/2$ case).}

{The eigenstates of the topological end state pairs are degenerate and can hybridize among themselves but the in-gap states with $Re[E]\ne 0$ are non-degenerate and positioned symmetrically about $E=0$ in the real energy axis\cite{edge2}. The amplitude distribution of the pair of end states indicate localization at the single end (one at left and the other at right) of the chain  deep within TNP[Fig.\ref{fig8}(b),(e)] but amplitude peaks at both the boundaries close to the TQPT point, be it in presence or absence of $\gamma$\cite{mandal,ep}. The in-gap states, however, always show localization at single end (decided by the sign of $\Delta/t$) of the chain. For $\theta=\pi/2$, a single pair of them exhibit peaks at one single boundary [see Fig.\ref{fig8}(c)] but for $\theta=\pi/4$, there are three pairs of in-gap states and all of them don't show peaks at same boundary [see Fig.\ref{fig8}(f)].  A summary of the characteristics of ZES and in-gap states for both the Hermitian and non-Hermitian system within TNP and TTP is given in Table.\ref{table:3}.}

{With increase on $\gamma$, the bulk states start having complex eigenvalues first at the TQPT $|\Delta/t|=\sqrt{2}$ for $\theta=\pi/2$. For $\theta=\pi/4$ we see two pairs of TQPTs and bulk states become complex at different $\gamma$ values for them, the minimum of that being registered at $|\Delta/t|=\sqrt{2(2-\sqrt{2})}$.}

%\red{The non-stationary states for $\gamma>\gamma_{ep}$ are characterized by instabilities like decay or growth with time\cite{zhu} and are important in designing various quantum computing and memory devices.}

{We should mention here that} asymmetric hopping in forward and backward directions in the SSH chain gives rise to the localization of bulk states near the boundary (in addition to the topological end states) and it can be understood as an outcome of the NH skin effect of non-reciprocal lattices {incorporating diagonal NH terms in the Hamiltonian\cite{yan,yuce,yao,martinez,song}. 
We can also mention here of a  study by Okuma $et.~al.$\cite{okuma} that refers to a} $\mathbb{Z}_{2}$ NH skin effect of a spinful canonical system which respects time-reversal symmetry i.e., $TT^{\star}=-1$. It reveals a pair of skin modes (Kramers doublet) localized at the two boundaries of the chain. 

\section{Summary and conclusions}
In this paper, we have considered a $\mathcal{PT}$ symmetric periodically hopping modulated NH SSH chain and investigated the topological properties of the same, particularly the behavior of its NH end, in-gap, and bulk states. This system possesses no chiral symmetry, instead, it respects a pseudo-anti-Hermiticity property. This work realizes the previously adopted connection\cite{wang} between TQPT and $\mathcal{SPT~BT}$ with more clarification. The $\mathcal{PT}$ symmetry of the whole system is determined by visualizing the symmetry property of individual end states\cite{pt}. Our analysis shows that within topological phase a $\mathcal{SPT~BT}$ occurs at EP where the end state eigenvalues coalesce to zero and then turn from real to imaginary. {In a finite chain,} the value of this $\gamma_{ep}$ is maximum close to the TQPT point within the TNP but if one looks deep into the TNP, one gets $\gamma_{ep}\rightarrow0$ resulting in the presence of $\mathcal{SPT~BP}$ (and absence of $\mathcal{PT}$-symmetric phase) for all arbitrary nonzero $\gamma$ values. On the other hand, our results in the TTP region are found consistent with that of Ref.\cite{zhu} which shows a $\mathcal{SPT~BT}$ at a $\gamma_{c}$ (considerably larger than $\gamma_{ep}$ found within TNP), below which the model has only real eigenvalues and vice-versa. Though it points at no such universal correlation between the topological properties and the $\mathcal{SPT~BT}$\cite{wang,zhu}, each region shows an interesting variation of $\mathcal{PT}$ symmetry based phases for all different hopping periodicities that we consider.
We have obtained the phase diagrams for $\theta=\pi$, and $\pi/2$ in $\gamma$-$\Delta$ plane {(both under PBC and OBC)} which show $\mathcal{PT}$ symmetric and partially and fully $\mathcal{PT}$ broken regions. The phase diagrams also reveal that the topological phase region has no dependency on $\gamma$ values which is also confirmed by the IPR plots {(Fig.\ref{ipr2}). However, in the maximally dimerized limit of $|\Delta/t|=1$, the eigenstates show an increase in localization with $\gamma$ (Fig.\ref{ipr3}). Though in general no skin effect is observed in a SSH chain with a diagonal NH term, such states localized at the end in addition to all the in-gap states can show some skin effect in this problem. The strength of IPRs gets lost gradually as one alters $\Delta$ from the TNP regime towards the TQPT point.}  Interestingly for $\theta=\pi/2$, the strength of IPRs for the two end states differ as there are different distributions of hopping periodicities as seen from the two ends. {In order to respect $\mathcal{PT}$ symmetry and extract $\gamma$ dependent winding number for our system, one can consider another form of winding number as studied in Ref.\cite{sudin} where winding number comes out to be unity (fraction or zero) for $\mathcal{PT}$ unbroken (partly broken or fully broken) phases.}
The end states of our NH system have been studied numerically.
The summary of the results obtained is enlisted in Table II and III.
%The $\mathcal{PT}$ symmetric phase transition point for weak $\gamma$ occurs at the same $\Delta$~(it is 0 for $\theta=\pi$) value at which the topological phase transition point is discernible for an isolated system.
%There is no such analytical expression to estimate EP for the systems having four sublattices. So it is worth studying to find the same.

{Our theoretical analysis of periodically hopping modulated NH SSH chain can instigate new applications like tuning phases in topological lasers or non-Hermitian superconductors\cite{prx9} or in topological light steering and funneling\cite{ep}. All these can be utilized in quantum computation purpose. Particularly, study of dynamics of such dissipative systems in presence of reservoirs at the edges can become very interesting and important in optical systems like lossy waveguides and cavities or for quantum information storage and processing purpose\cite{ep}. It is also worthwhile to extend these studies to bosonic systems in a cold-atom set up for the parameter variations there can be controlled rather easily\cite{pt}.
Lastly we should mention that in future we plan to look up more closely on how the wide variety of symmetries\cite{prx9} for a NH SSH chain, can be reconciled in different $\mathcal{PT}$ broken and/or unbroken phases of these periodically hopping modulated NH systems and how that can help in favorably modulating the loss/gain in dissipative open quantum systems.}
%Thus, the addition of $\gamma$ in our model doesn't show any noteworthy effect on the properties of all these eigenstates.
%Furthermore, the non-Hermitian end states can break the so-called $\mathcal{PT}$-symmetry by acquiring complex eigenvalues.

%In view of imaginary potential $U$, the DW/interface state has a vanishing real part of the energy, $Re[E]=0$ which supports the largest imaginary eigenvalues in the $(\gamma,Im[E])$ plane. Interestingly, the DW states for all commensurate $\theta$ values are free from the influence of non-Hermitian defects given at the center of the chain.

%\section*{Appendix}
\begin{widetext}
\begin{table*}
\parbox{0.94\linewidth}{
\centering
\begin{tabular}{  p{1.5cm}| p{2.5cm} |p{1cm}| p{7cm}|  p{1cm}|p{1cm}} 
\cline{1-6}
\multicolumn{1}{c}{} & \multicolumn{1}{c}{} & \multicolumn{1}{c}{}&\multicolumn{1}{c}{$\theta=\pi$}&\multicolumn{1}{c}{}\\ [0.65ex] 
 \hline
  System &   \multicolumn{1}{c}{} &\multicolumn{1}{c}{TNP}& &\multicolumn{1}{c}{TTP}\\ [1.25ex] 
  \cline{2-6}
   &  ZES & In-gap states&Properties&ZES&In-gap states \\ [1.25ex]
 \hline
 Hermitian  & Two zero energy states &$-$&Deep within the phase, these pair survive at a single boundary (one at the left and another at the right end of the chain) however, they can survive at both boundaries near TQPT.&$-$&$-$ \\ [1.2ex] 
 \hline
  Non-Hermitian &One imaginary conjugated pair. & $-$&ZES have equal share on both ends for $\gamma\le\gamma_{ep}$ when energies are real [Fig.\ref{fig3}(c)]. For $\gamma>\gamma_{ep}$, ZES becomes imaginary conjugated, showing increasingly larger peaks at single ends (one at left and the other at right) [Fig.\ref{fig3}(d)]. However, away from TQPT and $\gamma>\gamma_{ep}$, one ZES with imaginary energy shows a localization at left end while other at right boundary[Fig.\ref{fig2}(d)]. &$-$&$-$\\ [1.2ex] 
 \hline
\end{tabular}
\caption{Comparative features of Hermitian and non-Hermitian system for $\theta=\pi$.}
\label{table:2}}
\end{table*}
%\end{widetext}

%\begin{widetext}
\begin{table}
\parbox{1.0\linewidth}{
\centering
\begin{tabular}{  p{1.5cm}| p{2.0cm} |p{2.8cm}| p{5.9cm}|  p{1.1cm}| p{2cm}|  p{1.5cm}} 
\cline{1-7}
\multicolumn{1}{c}{} & \multicolumn{1}{c}{} & \multicolumn{1}{c}{}&\multicolumn{1}{c}{$\theta=\pi/2~ (\pi/4)$}&\multicolumn{1}{c}{}\\ [0.65ex] 
 \hline
  System &   \multicolumn{1}{c}{} &\multicolumn{1}{c}{TNP}& &\multicolumn{1}{c}{TTP}\\ [1.25ex] 
  \cline{2-7}
   &  ZES & In-gap states&Properties&ZES&In-gap states&Properties \\ [1.25ex]
 \hline
 Hermitian  & Two (Two) &Two (Six)& Deep within the phase, the pair of ZES survive at a single end. Near TQPT, they survive at both boundaries. In-gap states peak near the single end.&$-$&Two (Six)&Same as in TNP.\\ [1.2ex] 
 \hline
  Non-Hermitian &One imaginary conjugated pair (same)& One (three) pairs of purely real modes that become purely imaginary for large $\gamma$ and close to $\Delta/t=0$. &Only when $\gamma>\gamma_{ep}$, ZES becomes imaginary conjugated, showing peaks at both ends but with different amplitudes [Fig.\ref{fig8}(b)]. Away from TQPT, a single peak is observed at one end. The pairs of in-gap states always show peaks at single ends [Fig.\ref{fig8}(c)].&$-$&One complex pair (Three complex pairs)&same as in TNP. \\ [1.2ex] 
 \hline
\end{tabular}
\caption{Comparative features of Hermitian and non-Hermitian system for $\theta=\pi/2~(\pi/4)$.}
\label{table:3}}
\end{table}
\end{widetext}
%\newline

\section*{Acknowledgements}
%The authorsSK thanks B. Kumar, S. Basu, A. Saha and S. Mandal for fruitful discussions.
SM acknowledges the summer school TSACMP organized by IOP, Bhubaneswar that helped enriching the texts/explanations in the manuscript. SM and SK thank DST-SERB, Government of India for financially supporting this work via grant no. CRG/2022/002781.


\begin{thebibliography}{99}
\bibitem{comment}One can notice that the model has weak $\Delta$ ($\gamma$ will also be very small) close to the TQPT point.
\bibitem{comment2}The matrix $Q(k)$ is Hermitian i.e., $Q(k)^{\dagger}=Q(k)$ and it follows the anticommutation realtion $\{Q(k),\sigma_{3}\}=0$.
\bibitem{cmnt} For $\gamma\ne0$, we don't find any ZES (with absolute zero energy) and hence TQPT for such non-Hermitian case refers to the demarcation of presence and absence of NH ZES.
\end{thebibliography}
\end{document}